\documentclass[superscriptaddress, twocolumn, prl, showpacs, footinbib, a4]{revtex4-1}
\usepackage{graphicx}
\usepackage{enumerate}
\usepackage{tikz}
\usepackage{amsmath}
\usepackage{amssymb}
\usepackage{booktabs}
\usepackage{hyperref}
\usetikzlibrary{arrows,positioning}

\newcommand{\Lmin}{{L_\text{min}}}

\newcommand{\Lmax}{{L_\text{max}}}

\newcommand{\kpc}{\;\text{kpc}}
\newcommand{\gr}{$\gamma$-ray}

\newcommand{\ergs}{{\rm\, erg\,s^{-1}}}

\newcommand{\Fermi}{\textit{Fermi}}

\begin{document}

\title{Strong Support for the Millisecond Pulsar Origin of the Galactic Center
GeV Excess}

\author{Richard Bartels}
\email{r.t.bartels@uva.nl}

\author{Suraj Krishnamurthy}
\email{s.krishnamurthy@uva.nl}

\author{Christoph Weniger}
\email{c.weniger@uva.nl}
\affiliation{GRAPPA Institute, University of Amsterdam, Science Park 904, 1090
GL Amsterdam, Netherlands}

\date{4 February 2016}

\begin{abstract}
  Using \gr\ data from the \Fermi\ Large Area Telescope, various groups have
  identified a clear excess emission in the Inner Galaxy, at energies around a
  few GeV. This excess resembles remarkably well a signal from dark-matter
  annihilation. One of the most compelling astrophysical interpretations is
  that the excess is caused by the combined effect of a previously undetected
  population of dim \gr\ sources. Because of their spectral similarity, the best
  candidates are millisecond pulsars. Here, we search for this hypothetical
  source population, using a novel approach based on wavelet decomposition of
  the \gr\ sky and the statistics of Gaussian random fields.  Using almost
  seven years of \Fermi-LAT data, we detect a clustering of photons as
  predicted for the hypothetical population of millisecond pulsar, with a
  statistical significance of $10.0\sigma$.  For plausible values of the
  luminosity function, this population explains $100\%$ of the observed excess
  emission.  We argue that other extragalactic or Galactic sources, a
  mismodeling of Galactic diffuse emission, or the thick-disk population of
  pulsars are unlikely to account for this observation.  
\end{abstract}


\maketitle


\paragraph{Introduction.} Since its launch in 2008, the \Fermi\ Large Area
Telescope (LAT) has revolutionized our understanding of the \gr\ sky.  Among
the major successes are the detection of more than 3000 \gr\
sources~\cite{TheFermi-LAT:2015hja}, the discovery of the \Fermi\
bubbles~\cite{Su:2010qj}, some of the most stringent limits on dark-matter
annihilation~\cite{Ackermann:2015zua} and, most recently, the detection of
cross-correlations between the extragalactic \gr\ background and various galaxy
catalogs~\cite{Xia:2015wka}.

One of the most interesting \gr\ signatures identified in the \Fermi-LAT data
by various groups~\cite{Goodenough:2009gk, Vitale:2009hr, Hooper:2010mq,
Hooper:2011ti, Abazajian:2012pn, Gordon:2013vta, Macias:2013vya,
Abazajian:2014fta, Daylan:2014rsa, Zhou:2014lva, Calore:2014xka, TheFermi-LAT:2015kwa},
is an excess emission in the Inner Galaxy at energies around a few GeV.  This
excess attracted great attention because it has properties typical for a dark-matter 
annihilation signal.  This Galactic center excess (GCE) is
detected both within the inner 10 arcmin of the Galactic Center
(GC)~\cite{Hooper:2010mq, Abazajian:2012pn, Gordon:2013vta} and up to Galactic
latitudes of more than $10^\circ$~\cite{Hooper:2013rwa, Huang:2013pda,
Daylan:2014rsa, Calore:2014xka}.  It features a remarkably uniform spectrum and
approximately spherical symmetry~\cite{Daylan:2014rsa, Calore:2014xka}.
Proposed diffuse emission mechanisms, like leptonic or hadronic
outbursts~\cite{Carlson:2014cwa, Petrovic:2014uda, US} or cosmic-ray injection
in the central molecular zone~\cite{Gaggero:2015nsa}, potentially explain part
of the excess emission.  However, it is challenging to explain \emph{all} of
the above aspects of the GCE with these mechanisms alone.

Probably the most plausible astrophysical interpretation for the GCE is the
combined emission from a large number of unresolved millisecond pulsars (MSPs)
in the Galactic bulge region~\cite{Abazajian:2010zy, Abazajian:2014fta,
Gordon:2013vta, Yuan:2014rca}.  MSPs feature a spectrum compatible with the GCE
emission~\cite{Calore:2014xka}, and a large unresolved component can naturally
explain the uniformity of the GCE spectrum in different regions of the sky.
Recently, it was shown that the spatial distribution of MSPs that were spilled
out of disrupted globular clusters can explain the morphology of the
GCE~\cite{Brandt:2015ula}. Such MSPs from disrupted 
globular clusters have also been suggested as the source behind the
GeV through TeV emission in the inner
few parsec of the GC~\citep{Bednarek:2013oha}.
Further possible support for the
MSP hypothesis might come from \textit{Chandra} observations of low-mass x-ray
binaries (which are progenitor systems of MSPs) in M31, which show a centrally
peaked profile in the inner 2 kpc~\cite{Voss:2006az, 2012PhRvD..86h3511A}, as
well as the recent observation of extended hard X-ray emission from the
Galactic Center by \textit{NuSTAR}~\cite{2015Natur.520..646P}.

It was claimed that an interpretation of 100\% of the GCE emission in terms of
MSPs would be already ruled out: a sizeable fraction of the required
$10^3$--$10^4$ MSPs should have been already detected by the
\Fermi-LAT~\cite{Hooper:2013nhl, Cholis:2014lta}, but no (isolated) MSP has
been identified so far in the bulge region.  This conclusion depends crucially, 
however, on the adopted \gr\ luminosity of the brightest MSPs in the bulge
population, on the effective source sensitivity of \Fermi-LAT, and on the
treatment of unassociated sources in the Inner Galaxy~\cite{Petrovic:2014xra,
Brandt:2015ula}.  A realistic sensitivity study for MSPs in the context of the
GeV excess, taking into account all these effects, was lacking in the
literature up to now (but see Ref.~\cite{TheFermi-LAT:2013ssa}).

In this Letter, we close this gap and present a novel technique for the
analysis of dim \gr\ sources and apply it to \Fermi-LAT observations of the
Inner Galaxy.  Our method is based on the statistics of maxima in the
wavelet-transformed \gr\ sky (in context of \Fermi-LAT data,
wavelet transforms were used previously for the identification of point source
seeds~\cite{TheFermi-LAT:2015hja, TheFermi-LAT:2015kwa}). We search for contributions from a large
number of dim MSP-like sources, assuming that they are spatially distributed
as suggested by GCE observations.  Our method has several advantages
with respect to previously proposed techniques based on one-point
fluctuations~\cite{Lee:2014mza}, most notably the independence from Galactic
diffuse emission models and the ability for candidate source localization.

\medskip

\paragraph{Modeling.} We simulate a population of MSP-like 
sources, which we hereafter refer to simply as the central
source population (CSP), distributed around the GC at 8.5
kpc distance from the Sun. The CSP is taken to have a spatial distribution
that follows a radial power law with an index of $\Gamma=-2.5$ and a hard
cutoff at radius $r=3\kpc$~\cite{Daylan:2014rsa, Calore:2014xka}.  As a reference
\gr\ energy spectrum, we adopt the stacked MSP spectrum from
Ref.~\cite{Cholis:2014noa}, $\frac{dN}{dE} \propto e^{-E/{3.78\,\rm GeV}}
E^{-1.57}$.  The \gr\ luminosity function is modeled with a power law,
$\frac{dN}{dL}\propto L^{-\alpha}$, with index
$\alpha=-1.5$~\cite{Strong:2006hf, Cholis:2014noa, Venter:2014zea,
Petrovic:2014xra}, and with lower and upper
hard cutoffs at $\Lmin=10^{29} \ergs$ and $\Lmax=10^{34}$--$10^{36}\ergs$,
respectively.  Luminosities are integrated over 0.1--100 GeV.  Our results
depend little on $\Lmin$.  Given that only about 70 MSPs have been detected in
$\gamma$ rays up to now~\cite{TheFermi-LAT:2013ssa}, $\Lmax$ is not well constrained.
The \gr\ luminosity of the brightest observed MSP is somewhere in the range
($0.5$--$2)\times10^{35}\ergs$ \cite{TheFermi-LAT:2013ssa, Cholis:2014noa},
depending on the adopted source distance~\cite{Petrovic:2014xra,
Brandt:2015ula}.  Diffuse emission is modeled with the standard model for point
source analysis \texttt{gll\_iem\_v06.fits} and the corresponding isotropic
background.

\smallskip

\paragraph{Data.} For our analysis, we use almost seven years of ultraclean
\Fermi-LAT P8R2 data taken between August 4 2008 and June 3 2015 (we
find similar results for source class data). We select both
front- and back converted events in the energy range 1--4 GeV, which covers the
peak of the GCE spectrum.  The region of interest (ROI) covers the Inner Galaxy
and spans Galactic longitudes $|\ell| \leq 12^\circ$ and latitudes $2^\circ
\leq |b| \leq 12^\circ$.  The data are binned in Cartesian coordinates with a
pixel size of $0.1^\circ$.  

\smallskip

\paragraph{Wavelet peaks.} 
The wavelet transform of the \gr\ data is defined as the convolution of the
photon count map, $\mathcal{C}(\Omega)$, with the wavelet kernel,
$\mathcal{W}(\Omega)$,
\begin{equation}
    \mathcal{F}_\mathcal{W}[\mathcal{C}](\Omega) \equiv 
    \int d\Omega'\, \mathcal{W}(\Omega-\Omega') \mathcal{C}(\Omega')\;,
\end{equation}
where $\Omega$ denotes Galactic coordinates~\cite{Damiani:1997} [note that $\int d\Omega\,
\mathcal{W}(\Omega) = 0$].  The central observable for the current
analysis is the \emph{signal-to-noise ratio} (SNR) of the wavelet transform,
which we define as
\begin{equation}
    \mathcal{S}(\Omega) \equiv
    \frac{\mathcal{F}_\mathcal{W}[\mathcal{C}](\Omega)}
    {\sqrt{\mathcal{F}_{\mathcal{W}^2}[\mathcal{C}](\Omega)}}\;,
\end{equation}
where in the denominator the wavelet kernel is squared before performing the
convolution.  If the \gr\ flux varied only on scales much larger than the
extent of the wavelet kernel, and in the limit of a large number of photons,
$\mathcal{S}(\Omega)$ would behave like a smoothed Gaussian random field.
Consequentially, $\mathcal{S}(\Omega)$ can be loosely interpreted as the
local significance for having a source at position $\Omega$ in units of
standard deviations.

As the wavelet kernel, we adopt the second member of the mexican hat wavelet
family, which was shown to provide very good source discrimination
power~\cite{GonzalezNuevo:2006pa} and which was used for the identification of
compact sources in Planck data~\cite{Ade:2013jbr}.  The wavelet can be obtained
by a successive application of the Laplacian operator to a two-dimensional
Gaussian distribution with width $\sigma_b R$.  Here, $\sigma_b =
0.4^\circ$ corresponds to the \Fermi-LAT angular resolution at 1--4 GeV, and
$R$ is a tuning parameter.  We find best results when $R$ varies linearly with
latitude from $R=0.53$ at $b=0^\circ$ to $R=0.83$ at $b=\pm12^\circ$.  This
compensates to some degree the increasing diffuse backgrounds towards the
Galactic disk, while optimizing the source sensitivity at higher
latitudes~\cite{Ade:2013jbr}.

\begin{figure}
    \begin{center}
        \includegraphics[width=\linewidth]{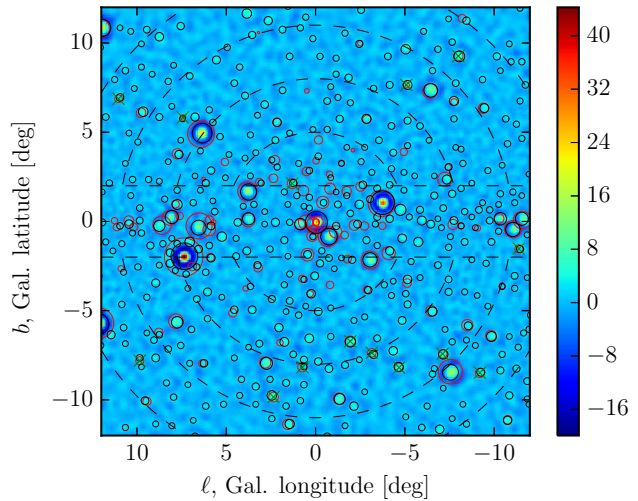}
    \end{center}
    \caption{SNR of the wavelet transform of $\gamma$ rays with energies in the range
      1--4 GeV, $\mathcal{S}(\Omega)$. The black circles show the position of
      wavelet peaks with $\mathcal{S}\geq2$; the red circles show the position
      of third Fermi-LAT catalog (3FGL) sources.  In both cases, the circle area scales with the
      significance of the source detection in that energy range.  The dashed
      lines indicate the regions that we use for the binned likelihood
      analysis, where latitudes $|b|<2^\circ$ are excluded because of the
      strong emission from the Galactic disk. The subset of 3FGL sources that
      remains unmasked in our analysis is indicated by the green crosses.}
    \label{fig:transform}
\end{figure}

The resulting SNR of the wavelet transform $\mathcal{S}(\Omega)$ is shown in
Fig.~\ref{fig:transform}.  As expected, the Galactic diffuse emission is almost
completely filtered out by the wavelet transform, whereas bright sources lead
to pronounced peaks.  We adopt a simple algorithm for peak identification: we
find all pixels in $\mathcal{S}(\Omega)$ with values larger than in the four
adjacent pixels.  We then clean these results from artifacts by forming
clusters of peaks with cophenetic distances less than $0.3^\circ$, and only
keep the most significant peak in each cluster. 

In Fig.~\ref{fig:transform}, we show the identified wavelet peaks with peak
significance $\mathcal{S}>2$, as well as all 3FGL sources for
comparison~\cite{TheFermi-LAT:2015hja}.  For sources that are bright enough in
the adopted energy range, we find a good correspondence between wavelet peaks
and the 3FGL, both in terms of position and significance (we compare the
significance of wavelet peaks $\mathcal{S}$ with the 1--3 GeV detection
significance for sources).

It is worth emphasizing that for the adopted spherically symmetric and
centrally peaked distribution of the CSP, most of the sources would be detected
not directly at the GC but a few degrees away from the Galactic disk.  This
is simply due to the much weaker diffuse emission at higher latitudes.  Our
focus on latitudes $|b|\geq 2^\circ$, thus, avoids regions where source detection
becomes less efficient, due to strong diffuse foregrounds, without significant
sensitivity loss for the source population of interest.

\smallskip

\paragraph{3FGL sources.} Before studying the statistics of the wavelet peaks
in detail below, we remove almost all peaks that correspond to the known 3FGL
sources based on a $0.3^\circ$ ($1^\circ$ for $\sqrt{TS}\geq50$) proximity cut.
However, in order to mitigate a potential bias on $\Lmax$, we do not mask peaks
that correspond to 3FGL sources that are likely part of the CSP.  We identify
such \emph{MSP candidate sources} by requiring that they (i) are tagged as
unassociated, (ii) show no indication for variability and (iii) have a spectrum
compatible with MSPs.  The last criterion is tested by performing a
$\chi^2$ fit of the above MSP reference spectrum to the spectrum given in the
3FGL (0.1--100 GeV; five energy bins).  Only the normalization is left free to
vary.  We require a fit quality of $\chi^2/\text{DOF}\leq 1.22$ (with DOF=4),
corresponding to a $p$ value $\geq0.3$.

We find 13 3FGL sources in the Inner Galaxy ROI that pass the above MSP cuts
(listed in the Supplemental Material).  Interestingly, the average number of
MSP candidate sources in same-sized control regions along the Galactic disk in
the range $|\ell| = 12^\circ $--$60^\circ$ is significantly smaller, with an
average of $3.1$.  It is tempting to interpret this excess of MSP candidate
sources in the Inner Galaxy as being caused by the brightest sources of the
CSP, above the less-pronounced thick-disk population of
MSPs~\cite{Gregoire:2013yta, Calore:2014oga}.  However, we emphasize that the status of these 13
sources is currently neither clear nor qualitatively decisive for our results.
Whether we mask them plays a minor role in the detection of the CSP below (but
it does affect the inferred values for $\Lmax$; see Supplemental Material).

\begin{figure}
    \begin{center}
        \includegraphics[width=\linewidth]{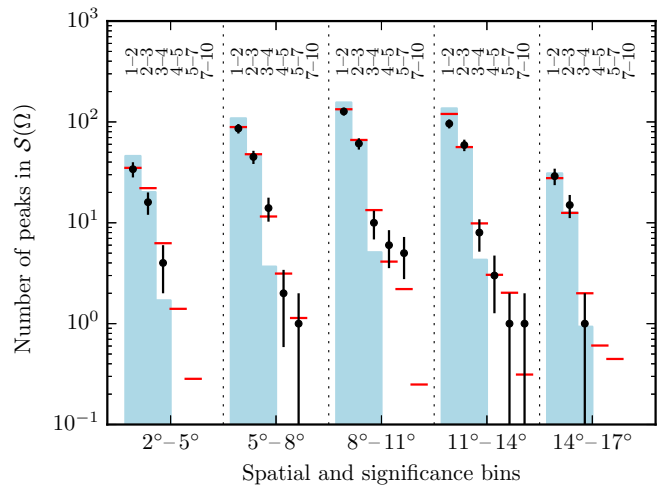}
    \end{center}
    \caption{Histogram of observed peaks in $\mathcal{S}(\Omega)$, in bins of
      projected radial distance from the GC and SNR values (black dots with
      statistical error bars).  We show the expectation value for the case of a
      negligible CSP as blue bars, whereas the expectation values for the
      best fit are shown in red.}
    \label{fig:hist}
\end{figure}

\smallskip

\paragraph{Statistical analysis.} In Fig.~\ref{fig:hist} we show a histogram of
the wavelet peaks in our ROI.  We bin the peaks in a two-dimensional grid,
which spans the projected angle from the Galactic Center $2^\circ$--$17^\circ$ and
wavelet peak significances in the range 1--10.  The bin edges are as indicated
in the figure.  As expected, photon shot noise gives rise to a large number of
peaks with low significances $\mathcal{S}\leq3$, and only a small number of
peaks has $\mathcal{S}\geq5$.

We assume that the number of peaks in each bin in Fig.~\ref{fig:hist} follows
-- in repeated experiments and random realizations of the CSP -- to a good
approximation a Poisson distribution.  We estimate the corresponding average
number of expected wavelet peaks in each bin using a large number of Monte
Carlo simulations, where we simulate the diffuse background emission,
random realizations of the source population and photon shot noise. 

In order to quantify what CSP luminosity function reproduces best the
observations, we perform a binned Poisson likelihood analysis of the wavelet
peak distribution.  The likelihood function is given by
\begin{equation}
    \mathcal{L} = \prod_{i=1}^{n_r}
    \prod_{j=1}^{n_s} \mathcal{P}(c_{ij}|\mu_{ij}(L_\text{max},
    \Phi_{5}))\;,
    \label{eqn:likelihood}
\end{equation}
where $n_r$ and $n_s$ are, respectively, the numbers of radial and peak SNR bins,
$c_{ij}$ is the observed and $\mu_{ij}$ the expected number of peaks, and
$\mathcal{P}$ is a Poisson distribution.  The expectation values depend
directly on the maximal luminosity, $\Lmax$, as well as on the number of
simulated sources, $n$.  To ease comparison with the literature, we determine
$n$ as a function of $\Phi_5$, which denotes the mean differential intensity of
the CSP at $b = \pm5^\circ$, $\ell=0^\circ$, and 2 GeV.  In the case of the GCE,
this value was found to be $\Phi_5^\text{GCE} = (8.5\pm 1.5)\times 10^{-7}\;\rm
GeV^{-1}\,cm^{-2}\,s^{-1}\,sr^{-1}$ at $95.4\%$ C.L.~\cite{Calore:2014xka}.

\begin{figure}
    \begin{center}
        \includegraphics[width=\linewidth]{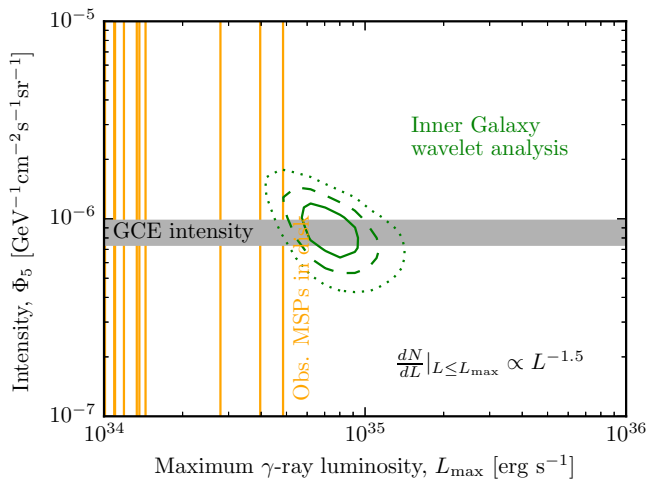}
    \end{center}
    \caption{Constraints on the maximum \gr\ luminosity of the CSP,
      $L_\text{max}$, and the population averaged intensity at $b=\pm5^\circ$,
      $\ell = 0^\circ$, and 2 GeV, $\Phi_5$, as derived from our wavelet
      analysis.  We show $68.7\%$, $95.4\%$, and $99.7\%$ C.L. contours.  We also
      indicate the values of $\Phi_5$ where the source population can explain
      $100\%$ of the GCE (horizontal gray band, $95.4\%$ CL), and as vertical
      orange lines the luminosity of the brightest observed nearby MSPs.}
    \label{fig:limits}
\end{figure}

\medskip

\paragraph{Results.} In Fig.~\ref{fig:hist}, we show the expectation values
that we obtain when neglecting contributions from the CSP (and any other
nondiffuse emission).  This corresponds to good approximation to the case
where the GCE is of truly diffuse origin, including the case of DM annihilation
or outburst events.  We find that the observed number of wavelet peaks with
$\mathcal{S}<2$ is significantly lower than expected, whereas the observed
number of peaks with $\mathcal{S}>3$ is significantly higher.  As we will show
next, this is precisely the effect that is caused by a dim source
population.

We now turn to the case with a nonzero CSP contribution.  In
Fig.~\ref{fig:limits}, we show the limits that we obtain on the two CSP
parameters when fitting the histogram in Fig.~\ref{fig:hist} as described
above.  We find that a nonzero contribution from the CSP is favored at the
level of at least $10.0\sigma$ (when quoting the statistical
significance, we conservatively take into account bins with $\mathcal{S}<5$
only, which are most affected by a dim source population, and least affected by
the masking of 3FGL sources [see the Supplemental Material for details]). The
best-fit value for the total differential intensity is $\Phi_5 =
(9.0\pm1.9)\times10^{-7}\ \rm GeV^{-1}cm^{-2}s^{-1}sr^{-1}$ and for the maximum
luminosity $\Lmax = (7.0\pm1.0)\times10^{34}\ \ergs$.  As can be seen in
Fig.~\ref{fig:hist}, we obtain in this case a very good fit to the data.  

Our preferred range of the maximum \gr\ luminosities reaches up to $\Lmax \leq
1.04\times10^{35}\ \ergs$ (at $95.4\%$ C.L), which is compatible with
observations of nearby MSPs.  We illustrate this by showing in
Fig.~\ref{fig:limits} the \gr\ luminosity of the brightest individually
observed nearby MSPs as given in Ref.~\cite{Cholis:2014noa} (we only
show objects where second \gr\ pulsar catalog~\cite{TheFermi-LAT:2013ssa} distances are available;
see Ref.~\cite{Brandt:2015ula} for a detailed discussion about distance
uncertainties). Furthermore, for the adopted slope of the luminosity function,
$\alpha=1.5$, the best-fit value for the total differential intensity of the
CSP $\Phi_5$ is consistent with the CSP accounting for 100\% of the GCE
emission.

\medskip

\paragraph{Discussion and conclusions.} We found corroborating evidence for the
hypothesis that the GCE is caused by a hitherto undetected population of
MSP-like sources.  We performed a wavelet transform of the \gr\ emission from
the Inner Galaxy, which removes Galactic diffuse emission and enhances point
sources, and we studied the statistics of the peaks in this transform.  We
detected with $10.0\sigma$ significance a suppression (enhancement) of low-
(high-) significance wavelet peaks, relative to the expectations for purely
diffuse emission.  We showed that this effect is caused by the presence of a
large number of dim point sources.  The spatial distribution of wavelet peaks
in the Inner Galaxy is compatible with a centrally peaked source distribution,
and the inferred cutoff of the \gr\ luminosity function of these sources agrees
with the observation of nearby MSPs.  This source population can, for
reasonable slopes of the luminosity function, account for $100\%$ of the GCE
emission.

For the purpose of this Letter, which introduces a novel technique, we kept
our analysis as simple as possible.  In general, one might worry that our results could be affected
by the presence of extragalactic and Galactic sources, by the thick-disk
population of MSPs and young pulsars, by the details of masking and unmasking
3FGL sources, by the details of the adopted \gr\ luminosity function, and by
unmodeled substructure in the Galactic diffuse emission that is not removed by the
wavelet transform.  We address all of these points in the Supplemental
Material
and show that it is rather unlikely that they affect our results
qualitatively, although quantitative changes in the obtained best-fit values
for $\Phi_5$ and $\Lmax$ are possible.  
In particular, we showed that the wavelet signal expected from the thick-disk population of
MSPs is an order of magnitude weaker than what we actually observed and that
interpretations related to unmodeled gas remain on closer inspection
unlikely.

The prospects for fully establishing the MSP interpretation within the coming
decade are very good.  Our results suggested that upcoming \gr\ observations with
improved angular resolution (planned or proposed \gr\ satellites like
GAMMA-400~\cite{gamma400}, ASTROGAM, and PANGU~\cite{pangu}) will allow us to
detect many more of the bulge sources and study their distribution and
spectra.  For current radio instruments, it remains rather
challenging to detect a MSP population in the bulge~\cite{Brandt:2015ula}, but
prospects for next-generation instruments are good.

\medskip

We thank for useful discussions with David Berge,
Francesca Calore, Ilias Cholis, Jennifer Gaskins, Dan Hooper, Simona Murgia,
Tracy Slatyer, Ben Safdi and Jacco Vink.
We thank the \Fermi\ Collaboration for providing the public \Fermi\ data as
well as the \Fermi\ Science Tools (\texttt{v10r0p5}).  SURFsara is thanked for
use of the Lisa Compute Cluster.  S.K.~and C.W.~are part of the VIDI
research programme ``Probing the Genesis of Dark Matter'', which is financed by
the Netherlands Organisation for Scientific Research (NWO).  R.B.~is part of a
GRAPPA-Ph.D.~program funded by NWO.  Part of this work was performed at the Aspen
Center for Physics, which is supported by National Science Foundation Grant No. PHY-1066293.

\medskip

\textit{Note added in proof.}--Recently, we became aware of
another group studying dim \gr\ sources in the Inner
Galaxy, using non-Poissonian photon statistics~\cite{Lee:2015fea}.

\bibliographystyle{apsrev4-1}
\bibliography{waveletGC}

\begin{thebibliography}{49}%
\makeatletter
\providecommand \@ifxundefined [1]{%
 \@ifx{#1\undefined}
}%
\providecommand \@ifnum [1]{%
 \ifnum #1\expandafter \@firstoftwo
 \else \expandafter \@secondoftwo
 \fi
}%
\providecommand \@ifx [1]{%
 \ifx #1\expandafter \@firstoftwo
 \else \expandafter \@secondoftwo
 \fi
}%
\providecommand \natexlab [1]{#1}%
\providecommand \enquote  [1]{``#1''}%
\providecommand \bibnamefont  [1]{#1}%
\providecommand \bibfnamefont [1]{#1}%
\providecommand \citenamefont [1]{#1}%
\providecommand \href@noop [0]{\@secondoftwo}%
\providecommand \href [0]{\begingroup \@sanitize@url \@href}%
\providecommand \@href[1]{\@@startlink{#1}\@@href}%
\providecommand \@@href[1]{\endgroup#1\@@endlink}%
\providecommand \@sanitize@url [0]{\catcode `\\12\catcode `\$12\catcode
  `\&12\catcode `\#12\catcode `\^12\catcode `\_12\catcode `\%12\relax}%
\providecommand \@@startlink[1]{}%
\providecommand \@@endlink[0]{}%
\providecommand \url  [0]{\begingroup\@sanitize@url \@url }%
\providecommand \@url [1]{\endgroup\@href {#1}{\urlprefix }}%
\providecommand \urlprefix  [0]{URL }%
\providecommand \Eprint [0]{\href }%
\providecommand \doibase [0]{http://dx.doi.org/}%
\providecommand \selectlanguage [0]{\@gobble}%
\providecommand \bibinfo  [0]{\@secondoftwo}%
\providecommand \bibfield  [0]{\@secondoftwo}%
\providecommand \translation [1]{[#1]}%
\providecommand \BibitemOpen [0]{}%
\providecommand \bibitemStop [0]{}%
\providecommand \bibitemNoStop [0]{.\EOS\space}%
\providecommand \EOS [0]{\spacefactor3000\relax}%
\providecommand \BibitemShut  [1]{\csname bibitem#1\endcsname}%
\let\auto@bib@innerbib\@empty
\bibitem [{\citenamefont {Acero}\ \emph {et~al.}(2015)\citenamefont {Acero}
  \emph {et~al.}}]{TheFermi-LAT:2015hja}%
  \BibitemOpen
  \bibfield  {author} {\bibinfo {author} {\bibfnamefont {F.}~\bibnamefont
  {Acero}} \emph {et~al.} (\bibinfo {collaboration} {Fermi-LAT
  Collaboration}),\ }\href {\doibase 10.1088/0067-0049/218/2/23} {\bibfield
  {journal} {\bibinfo  {journal} {Astrophys. J. Suppl.}\ }\textbf {\bibinfo
  {volume} {218}},\ \bibinfo {pages} {23} (\bibinfo {year} {2015})}\BibitemShut
  {NoStop}%
\bibitem [{\citenamefont {Su}\ \emph {et~al.}(2010)\citenamefont {Su},
  \citenamefont {Slatyer},\ and\ \citenamefont {Finkbeiner}}]{Su:2010qj}%
  \BibitemOpen
  \bibfield  {author} {\bibinfo {author} {\bibfnamefont {M.}~\bibnamefont
  {Su}}, \bibinfo {author} {\bibfnamefont {T.~R.}\ \bibnamefont {Slatyer}}, \
  and\ \bibinfo {author} {\bibfnamefont {D.~P.}\ \bibnamefont {Finkbeiner}},\
  }\href {\doibase 10.1088/0004-637X/724/2/1044} {\bibfield  {journal}
  {\bibinfo  {journal} {Astrophys. J.}\ }\textbf {\bibinfo {volume} {724}},\
  \bibinfo {pages} {1044} (\bibinfo {year} {2010})},\ \Eprint
  {http://arxiv.org/abs/1005.5480} {arXiv:1005.5480 [astro-ph.HE]} \BibitemShut
  {NoStop}%
\bibitem [{\citenamefont {Ackermann}\ \emph {et~al.}(2015)\citenamefont
  {Ackermann} \emph {et~al.}}]{Ackermann:2015zua}%
  \BibitemOpen
  \bibfield  {author} {\bibinfo {author} {\bibfnamefont {M.}~\bibnamefont
  {Ackermann}} \emph {et~al.} (\bibinfo {collaboration} {Fermi-LAT
  Collaboration}),\ }\href {\doibase 10.1103/PhysRevLett.115.231301} {\bibfield
   {journal} {\bibinfo  {journal} {Phys. Rev. Lett.}\ }\textbf {\bibinfo
  {volume} {115}},\ \bibinfo {pages} {231301} (\bibinfo {year} {2015})},\
  \Eprint {http://arxiv.org/abs/1503.02641} {arXiv:1503.02641 [astro-ph.HE]}
  \BibitemShut {NoStop}%
\bibitem [{\citenamefont {Xia}\ \emph {et~al.}(2015)\citenamefont {Xia},
  \citenamefont {Cuoco}, \citenamefont {Branchini},\ and\ \citenamefont
  {Viel}}]{Xia:2015wka}%
  \BibitemOpen
  \bibfield  {author} {\bibinfo {author} {\bibfnamefont {J.-Q.}\ \bibnamefont
  {Xia}}, \bibinfo {author} {\bibfnamefont {A.}~\bibnamefont {Cuoco}}, \bibinfo
  {author} {\bibfnamefont {E.}~\bibnamefont {Branchini}}, \ and\ \bibinfo
  {author} {\bibfnamefont {M.}~\bibnamefont {Viel}},\ }\href {\doibase
  10.1088/0067-0049/217/1/15} {\bibfield  {journal} {\bibinfo  {journal}
  {Astrophys.J.Suppl.}\ }\textbf {\bibinfo {volume} {217}},\ \bibinfo {pages}
  {15} (\bibinfo {year} {2015})},\ \Eprint {http://arxiv.org/abs/1503.05918}
  {arXiv:1503.05918 [astro-ph.CO]} \BibitemShut {NoStop}%
\bibitem [{\citenamefont {Goodenough}\ and\ \citenamefont
  {Hooper}(2009)}]{Goodenough:2009gk}%
  \BibitemOpen
  \bibfield  {author} {\bibinfo {author} {\bibfnamefont {L.}~\bibnamefont
  {Goodenough}}\ and\ \bibinfo {author} {\bibfnamefont {D.}~\bibnamefont
  {Hooper}},\ }\href@noop {} {\  (\bibinfo {year} {2009})},\ \Eprint
  {http://arxiv.org/abs/0910.2998} {arXiv:0910.2998 [hep-ph]} \BibitemShut
  {NoStop}%
\bibitem [{\citenamefont {Vitale}\ and\ \citenamefont
  {Morselli}(2009)}]{Vitale:2009hr}%
  \BibitemOpen
  \bibfield  {author} {\bibinfo {author} {\bibfnamefont {V.}~\bibnamefont
  {Vitale}}\ and\ \bibinfo {author} {\bibfnamefont {A.}~\bibnamefont
  {Morselli}} (\bibinfo {collaboration} {Fermi-LAT Collaboration}),\
  }\href@noop {} {\  (\bibinfo {year} {2009})},\ \Eprint
  {http://arxiv.org/abs/0912.3828} {arXiv:0912.3828 [astro-ph.HE]} \BibitemShut
  {NoStop}%
\bibitem [{\citenamefont {Hooper}\ and\ \citenamefont
  {Goodenough}(2011)}]{Hooper:2010mq}%
  \BibitemOpen
  \bibfield  {author} {\bibinfo {author} {\bibfnamefont {D.}~\bibnamefont
  {Hooper}}\ and\ \bibinfo {author} {\bibfnamefont {L.}~\bibnamefont
  {Goodenough}},\ }\href {\doibase 10.1016/j.physletb.2011.02.029} {\bibfield
  {journal} {\bibinfo  {journal} {Phys.Lett.}\ }\textbf {\bibinfo {volume}
  {B697}},\ \bibinfo {pages} {412} (\bibinfo {year} {2011})},\ \Eprint
  {http://arxiv.org/abs/1010.2752} {arXiv:1010.2752 [hep-ph]} \BibitemShut
  {NoStop}%
\bibitem [{\citenamefont {Hooper}\ and\ \citenamefont
  {Linden}(2011)}]{Hooper:2011ti}%
  \BibitemOpen
  \bibfield  {author} {\bibinfo {author} {\bibfnamefont {D.}~\bibnamefont
  {Hooper}}\ and\ \bibinfo {author} {\bibfnamefont {T.}~\bibnamefont
  {Linden}},\ }\href {\doibase 10.1103/PhysRevD.84.123005} {\bibfield
  {journal} {\bibinfo  {journal} {Phys.Rev.}\ }\textbf {\bibinfo {volume}
  {D84}},\ \bibinfo {pages} {123005} (\bibinfo {year} {2011})},\ \Eprint
  {http://arxiv.org/abs/1110.0006} {arXiv:1110.0006 [astro-ph.HE]} \BibitemShut
  {NoStop}%
\bibitem [{\citenamefont {Abazajian}\ and\ \citenamefont
  {Kaplinghat}(2012)}]{Abazajian:2012pn}%
  \BibitemOpen
  \bibfield  {author} {\bibinfo {author} {\bibfnamefont {K.~N.}\ \bibnamefont
  {Abazajian}}\ and\ \bibinfo {author} {\bibfnamefont {M.}~\bibnamefont
  {Kaplinghat}},\ }\href {\doibase 10.1103/PhysRevD.86.083511} {\bibfield
  {journal} {\bibinfo  {journal} {Phys.Rev.}\ }\textbf {\bibinfo {volume}
  {D86}},\ \bibinfo {pages} {083511} (\bibinfo {year} {2012})},\ \Eprint
  {http://arxiv.org/abs/1207.6047} {arXiv:1207.6047 [astro-ph.HE]} \BibitemShut
  {NoStop}%
\bibitem [{\citenamefont {Gordon}\ and\ \citenamefont
  {Macias}(2013)}]{Gordon:2013vta}%
  \BibitemOpen
  \bibfield  {author} {\bibinfo {author} {\bibfnamefont {C.}~\bibnamefont
  {Gordon}}\ and\ \bibinfo {author} {\bibfnamefont {O.}~\bibnamefont
  {Macias}},\ }\href {\doibase 10.1103/PhysRevD.88.083521} {\bibfield
  {journal} {\bibinfo  {journal} {Phys.Rev.}\ }\textbf {\bibinfo {volume}
  {D88}},\ \bibinfo {pages} {083521} (\bibinfo {year} {2013})},\ \Eprint
  {http://arxiv.org/abs/1306.5725} {arXiv:1306.5725 [astro-ph.HE]} \BibitemShut
  {NoStop}%
\bibitem [{\citenamefont {Macias}\ and\ \citenamefont
  {Gordon}(2014)}]{Macias:2013vya}%
  \BibitemOpen
  \bibfield  {author} {\bibinfo {author} {\bibfnamefont {O.}~\bibnamefont
  {Macias}}\ and\ \bibinfo {author} {\bibfnamefont {C.}~\bibnamefont
  {Gordon}},\ }\href {\doibase 10.1103/PhysRevD.89.063515} {\bibfield
  {journal} {\bibinfo  {journal} {Phys.Rev.}\ }\textbf {\bibinfo {volume}
  {D89}},\ \bibinfo {pages} {063515} (\bibinfo {year} {2014})},\ \Eprint
  {http://arxiv.org/abs/1312.6671} {arXiv:1312.6671 [astro-ph.HE]} \BibitemShut
  {NoStop}%
\bibitem [{\citenamefont {Abazajian}\ \emph {et~al.}(2014)\citenamefont
  {Abazajian}, \citenamefont {Canac}, \citenamefont {Horiuchi},\ and\
  \citenamefont {Kaplinghat}}]{Abazajian:2014fta}%
  \BibitemOpen
  \bibfield  {author} {\bibinfo {author} {\bibfnamefont {K.~N.}\ \bibnamefont
  {Abazajian}}, \bibinfo {author} {\bibfnamefont {N.}~\bibnamefont {Canac}},
  \bibinfo {author} {\bibfnamefont {S.}~\bibnamefont {Horiuchi}}, \ and\
  \bibinfo {author} {\bibfnamefont {M.}~\bibnamefont {Kaplinghat}},\ }\href
  {\doibase 10.1103/PhysRevD.90.023526} {\bibfield  {journal} {\bibinfo
  {journal} {Phys.Rev.}\ }\textbf {\bibinfo {volume} {D90}},\ \bibinfo {pages}
  {023526} (\bibinfo {year} {2014})},\ \Eprint {http://arxiv.org/abs/1402.4090}
  {arXiv:1402.4090 [astro-ph.HE]} \BibitemShut {NoStop}%
\bibitem [{\citenamefont {Daylan}\ \emph {et~al.}(2014)\citenamefont {Daylan},
  \citenamefont {Finkbeiner}, \citenamefont {Hooper}, \citenamefont {Linden},
  \citenamefont {Portillo} \emph {et~al.}}]{Daylan:2014rsa}%
  \BibitemOpen
  \bibfield  {author} {\bibinfo {author} {\bibfnamefont {T.}~\bibnamefont
  {Daylan}}, \bibinfo {author} {\bibfnamefont {D.~P.}\ \bibnamefont
  {Finkbeiner}}, \bibinfo {author} {\bibfnamefont {D.}~\bibnamefont {Hooper}},
  \bibinfo {author} {\bibfnamefont {T.}~\bibnamefont {Linden}}, \bibinfo
  {author} {\bibfnamefont {S.~K.~N.}\ \bibnamefont {Portillo}},  \emph
  {et~al.},\ }\href@noop {} {\  (\bibinfo {year} {2014})},\ \Eprint
  {http://arxiv.org/abs/1402.6703} {arXiv:1402.6703 [astro-ph.HE]} \BibitemShut
  {NoStop}%
\bibitem [{\citenamefont {Zhou}\ \emph {et~al.}(2015)\citenamefont {Zhou},
  \citenamefont {Liang}, \citenamefont {Huang}, \citenamefont {Li},
  \citenamefont {Fan}, \citenamefont {Feng},\ and\ \citenamefont
  {Chang}}]{Zhou:2014lva}%
  \BibitemOpen
  \bibfield  {author} {\bibinfo {author} {\bibfnamefont {B.}~\bibnamefont
  {Zhou}}, \bibinfo {author} {\bibfnamefont {Y.-F.}\ \bibnamefont {Liang}},
  \bibinfo {author} {\bibfnamefont {X.}~\bibnamefont {Huang}}, \bibinfo
  {author} {\bibfnamefont {X.}~\bibnamefont {Li}}, \bibinfo {author}
  {\bibfnamefont {Y.-Z.}\ \bibnamefont {Fan}}, \bibinfo {author} {\bibfnamefont
  {L.}~\bibnamefont {Feng}}, \ and\ \bibinfo {author} {\bibfnamefont
  {J.}~\bibnamefont {Chang}},\ }\href {\doibase 10.1103/PhysRevD.91.123010}
  {\bibfield  {journal} {\bibinfo  {journal} {Phys. Rev.}\ }\textbf {\bibinfo
  {volume} {D91}},\ \bibinfo {pages} {123010} (\bibinfo {year} {2015})},\
  \Eprint {http://arxiv.org/abs/1406.6948} {arXiv:1406.6948 [astro-ph.HE]}
  \BibitemShut {NoStop}%
\bibitem [{\citenamefont {Calore}\ \emph {et~al.}(2015)\citenamefont {Calore},
  \citenamefont {Cholis},\ and\ \citenamefont {Weniger}}]{Calore:2014xka}%
  \BibitemOpen
  \bibfield  {author} {\bibinfo {author} {\bibfnamefont {F.}~\bibnamefont
  {Calore}}, \bibinfo {author} {\bibfnamefont {I.}~\bibnamefont {Cholis}}, \
  and\ \bibinfo {author} {\bibfnamefont {C.}~\bibnamefont {Weniger}},\ }\href
  {\doibase 10.1088/1475-7516/2015/03/038} {\bibfield  {journal} {\bibinfo
  {journal} {JCAP}\ }\textbf {\bibinfo {volume} {1503}},\ \bibinfo {pages}
  {038} (\bibinfo {year} {2015})},\ \Eprint {http://arxiv.org/abs/1409.0042}
  {arXiv:1409.0042 [astro-ph.CO]} \BibitemShut {NoStop}%
\bibitem [{\citenamefont {Ajello}\ \emph {et~al.}(2015)\citenamefont {Ajello}
  \emph {et~al.}}]{TheFermi-LAT:2015kwa}%
  \BibitemOpen
  \bibfield  {author} {\bibinfo {author} {\bibfnamefont {M.}~\bibnamefont
  {Ajello}} \emph {et~al.} (\bibinfo {collaboration} {Fermi-LAT
  Collaboration}),\ }\href@noop {} {\  (\bibinfo {year} {2015})},\ \Eprint
  {http://arxiv.org/abs/1511.02938} {arXiv:1511.02938 [astro-ph.HE]}
  \BibitemShut {NoStop}%
\bibitem [{\citenamefont {Hooper}\ and\ \citenamefont
  {Slatyer}(2013)}]{Hooper:2013rwa}%
  \BibitemOpen
  \bibfield  {author} {\bibinfo {author} {\bibfnamefont {D.}~\bibnamefont
  {Hooper}}\ and\ \bibinfo {author} {\bibfnamefont {T.~R.}\ \bibnamefont
  {Slatyer}},\ }\href {\doibase 10.1016/j.dark.2013.06.003} {\bibfield
  {journal} {\bibinfo  {journal} {Phys.Dark Univ.}\ }\textbf {\bibinfo {volume}
  {2}},\ \bibinfo {pages} {118} (\bibinfo {year} {2013})},\ \Eprint
  {http://arxiv.org/abs/1302.6589} {arXiv:1302.6589 [astro-ph.HE]} \BibitemShut
  {NoStop}%
\bibitem [{\citenamefont {Huang}\ \emph {et~al.}(2013)\citenamefont {Huang},
  \citenamefont {Urbano},\ and\ \citenamefont {Xue}}]{Huang:2013pda}%
  \BibitemOpen
  \bibfield  {author} {\bibinfo {author} {\bibfnamefont {W.-C.}\ \bibnamefont
  {Huang}}, \bibinfo {author} {\bibfnamefont {A.}~\bibnamefont {Urbano}}, \
  and\ \bibinfo {author} {\bibfnamefont {W.}~\bibnamefont {Xue}},\ }\href@noop
  {} {\  (\bibinfo {year} {2013})},\ \Eprint {http://arxiv.org/abs/1307.6862}
  {arXiv:1307.6862 [hep-ph]} \BibitemShut {NoStop}%
\bibitem [{\citenamefont {Carlson}\ and\ \citenamefont
  {Profumo}(2014)}]{Carlson:2014cwa}%
  \BibitemOpen
  \bibfield  {author} {\bibinfo {author} {\bibfnamefont {E.}~\bibnamefont
  {Carlson}}\ and\ \bibinfo {author} {\bibfnamefont {S.}~\bibnamefont
  {Profumo}},\ }\href {\doibase 10.1103/PhysRevD.90.023015} {\bibfield
  {journal} {\bibinfo  {journal} {Phys.Rev.}\ }\textbf {\bibinfo {volume}
  {D90}},\ \bibinfo {pages} {023015} (\bibinfo {year} {2014})},\ \Eprint
  {http://arxiv.org/abs/1405.7685} {arXiv:1405.7685 [astro-ph.HE]} \BibitemShut
  {NoStop}%
\bibitem [{\citenamefont {Petrovic}\ \emph {et~al.}(2014)\citenamefont
  {Petrovic}, \citenamefont {Serpico},\ and\ \citenamefont
  {Zaharijas}}]{Petrovic:2014uda}%
  \BibitemOpen
  \bibfield  {author} {\bibinfo {author} {\bibfnamefont {J.}~\bibnamefont
  {Petrovic}}, \bibinfo {author} {\bibfnamefont {P.~D.}\ \bibnamefont
  {Serpico}}, \ and\ \bibinfo {author} {\bibfnamefont {G.}~\bibnamefont
  {Zaharijas}},\ }\href {\doibase 10.1088/1475-7516/2014/10/052} {\bibfield
  {journal} {\bibinfo  {journal} {JCAP}\ }\textbf {\bibinfo {volume} {1410}},\
  \bibinfo {pages} {052} (\bibinfo {year} {2014})},\ \Eprint
  {http://arxiv.org/abs/1405.7928} {arXiv:1405.7928 [astro-ph.HE]} \BibitemShut
  {NoStop}%
\bibitem [{\citenamefont {Cholis}\ \emph
  {et~al.}(2015{\natexlab{a}})\citenamefont {Cholis}, \citenamefont {Evoli},
  \citenamefont {Calore}, \citenamefont {Linden}, \citenamefont {Weniger},\
  and\ \citenamefont {Hooper}}]{US}%
  \BibitemOpen
  \bibfield  {author} {\bibinfo {author} {\bibfnamefont {I.}~\bibnamefont
  {Cholis}}, \bibinfo {author} {\bibfnamefont {C.}~\bibnamefont {Evoli}},
  \bibinfo {author} {\bibfnamefont {F.}~\bibnamefont {Calore}}, \bibinfo
  {author} {\bibfnamefont {T.}~\bibnamefont {Linden}}, \bibinfo {author}
  {\bibfnamefont {C.}~\bibnamefont {Weniger}}, \ and\ \bibinfo {author}
  {\bibfnamefont {D.}~\bibnamefont {Hooper}},\ }\href@noop {} {\  (\bibinfo
  {year} {2015}{\natexlab{a}})},\ \Eprint {http://arxiv.org/abs/1506.05119}
  {arXiv:1506.05119 [astro-ph.HE]} \BibitemShut {NoStop}%
\bibitem [{\citenamefont {Gaggero}\ \emph {et~al.}(2015)\citenamefont
  {Gaggero}, \citenamefont {Taoso}, \citenamefont {Urbano}, \citenamefont
  {Valli},\ and\ \citenamefont {Ullio}}]{Gaggero:2015nsa}%
  \BibitemOpen
  \bibfield  {author} {\bibinfo {author} {\bibfnamefont {D.}~\bibnamefont
  {Gaggero}}, \bibinfo {author} {\bibfnamefont {M.}~\bibnamefont {Taoso}},
  \bibinfo {author} {\bibfnamefont {A.}~\bibnamefont {Urbano}}, \bibinfo
  {author} {\bibfnamefont {M.}~\bibnamefont {Valli}}, \ and\ \bibinfo {author}
  {\bibfnamefont {P.}~\bibnamefont {Ullio}},\ }\href@noop {} {\  (\bibinfo
  {year} {2015})},\ \Eprint {http://arxiv.org/abs/1507.06129} {arXiv:1507.06129
  [astro-ph.HE]} \BibitemShut {NoStop}%
\bibitem [{\citenamefont {Abazajian}(2011)}]{Abazajian:2010zy}%
  \BibitemOpen
  \bibfield  {author} {\bibinfo {author} {\bibfnamefont {K.~N.}\ \bibnamefont
  {Abazajian}},\ }\href {\doibase 10.1088/1475-7516/2011/03/010} {\bibfield
  {journal} {\bibinfo  {journal} {JCAP}\ }\textbf {\bibinfo {volume} {1103}},\
  \bibinfo {pages} {010} (\bibinfo {year} {2011})},\ \Eprint
  {http://arxiv.org/abs/1011.4275} {arXiv:1011.4275 [astro-ph.HE]} \BibitemShut
  {NoStop}%
\bibitem [{\citenamefont {Yuan}\ and\ \citenamefont
  {Zhang}(2014)}]{Yuan:2014rca}%
  \BibitemOpen
  \bibfield  {author} {\bibinfo {author} {\bibfnamefont {Q.}~\bibnamefont
  {Yuan}}\ and\ \bibinfo {author} {\bibfnamefont {B.}~\bibnamefont {Zhang}},\
  }\href {\doibase 10.1016/j.jheap.2014.06.001} {\bibfield  {journal} {\bibinfo
   {journal} {JHEAp}\ }\textbf {\bibinfo {volume} {3-4}},\ \bibinfo {pages} {1}
  (\bibinfo {year} {2014})},\ \Eprint {http://arxiv.org/abs/1404.2318}
  {arXiv:1404.2318 [astro-ph.HE]} \BibitemShut {NoStop}%
\bibitem [{\citenamefont {Brandt}\ and\ \citenamefont
  {Kocsis}(2015)}]{Brandt:2015ula}%
  \BibitemOpen
  \bibfield  {author} {\bibinfo {author} {\bibfnamefont {T.~D.}\ \bibnamefont
  {Brandt}}\ and\ \bibinfo {author} {\bibfnamefont {B.}~\bibnamefont
  {Kocsis}},\ }\href {\doibase 10.1088/0004-637X/812/1/15} {\bibfield
  {journal} {\bibinfo  {journal} {Astrophys. J.}\ }\textbf {\bibinfo {volume}
  {812}},\ \bibinfo {pages} {15} (\bibinfo {year} {2015})},\ \Eprint
  {http://arxiv.org/abs/1507.05616} {arXiv:1507.05616 [astro-ph.HE]}
  \BibitemShut {NoStop}%
\bibitem [{\citenamefont {Bednarek}\ and\ \citenamefont
  {Sobczak}(2013)}]{Bednarek:2013oha}%
  \BibitemOpen
  \bibfield  {author} {\bibinfo {author} {\bibfnamefont {W.}~\bibnamefont
  {Bednarek}}\ and\ \bibinfo {author} {\bibfnamefont {T.}~\bibnamefont
  {Sobczak}},\ }\href {\doibase 10.1093/mnrasl/slt084} {\bibfield  {journal}
  {\bibinfo  {journal} {Mon. Not. Roy. Astron. Soc.}\ }\textbf {\bibinfo
  {volume} {435}},\ \bibinfo {pages} {L14} (\bibinfo {year} {2013})},\ \Eprint
  {http://arxiv.org/abs/1306.4760} {arXiv:1306.4760 [astro-ph.HE]} \BibitemShut
  {NoStop}%
\bibitem [{\citenamefont {{Voss}}\ and\ \citenamefont
  {{Gilfanov}}(2007)}]{Voss:2006az}%
  \BibitemOpen
  \bibfield  {author} {\bibinfo {author} {\bibfnamefont {R.}~\bibnamefont
  {{Voss}}}\ and\ \bibinfo {author} {\bibfnamefont {M.}~\bibnamefont
  {{Gilfanov}}},\ }\href {\doibase 10.1051/0004-6361:20066614} {\bibfield
  {journal} {\bibinfo  {journal} {Astron. and Astroph.}\ }\textbf {\bibinfo
  {volume} {468}},\ \bibinfo {pages} {49} (\bibinfo {year} {2007})},\ \Eprint
  {http://arxiv.org/abs/astro-ph/0610649} {astro-ph/0610649} \BibitemShut
  {NoStop}%
\bibitem [{\citenamefont {{Abazajian}}\ and\ \citenamefont
  {{Kaplinghat}}(2012)}]{2012PhRvD..86h3511A}%
  \BibitemOpen
  \bibfield  {author} {\bibinfo {author} {\bibfnamefont {K.~N.}\ \bibnamefont
  {{Abazajian}}}\ and\ \bibinfo {author} {\bibfnamefont {M.}~\bibnamefont
  {{Kaplinghat}}},\ }\href {\doibase 10.1103/PhysRevD.86.083511} {\bibfield
  {journal} {\bibinfo  {journal} {\prd}\ }\textbf {\bibinfo {volume} {86}},\
  \bibinfo {eid} {083511} (\bibinfo {year} {2012})},\ \Eprint
  {http://arxiv.org/abs/1207.6047} {arXiv:1207.6047 [astro-ph.HE]} \BibitemShut
  {NoStop}%
\bibitem [{\citenamefont {{Perez}}\ \emph {et~al.}(2015)\citenamefont {{Perez}}
  \emph {et~al.}}]{2015Natur.520..646P}%
  \BibitemOpen
  \bibfield  {author} {\bibinfo {author} {\bibfnamefont {K.}~\bibnamefont
  {{Perez}}} \emph {et~al.},\ }\href {\doibase 10.1038/nature14353} {\bibfield
  {journal} {\bibinfo  {journal} {\nat}\ }\textbf {\bibinfo {volume} {520}},\
  \bibinfo {pages} {646} (\bibinfo {year} {2015})}\BibitemShut {NoStop}%
\bibitem [{\citenamefont {Hooper}\ \emph {et~al.}(2013)\citenamefont {Hooper},
  \citenamefont {Cholis}, \citenamefont {Linden}, \citenamefont
  {Siegal-Gaskins},\ and\ \citenamefont {Slatyer}}]{Hooper:2013nhl}%
  \BibitemOpen
  \bibfield  {author} {\bibinfo {author} {\bibfnamefont {D.}~\bibnamefont
  {Hooper}}, \bibinfo {author} {\bibfnamefont {I.}~\bibnamefont {Cholis}},
  \bibinfo {author} {\bibfnamefont {T.}~\bibnamefont {Linden}}, \bibinfo
  {author} {\bibfnamefont {J.~M.}\ \bibnamefont {Siegal-Gaskins}}, \ and\
  \bibinfo {author} {\bibfnamefont {T.~R.}\ \bibnamefont {Slatyer}},\ }\href
  {\doibase 10.1103/PhysRevD.88.083009} {\bibfield  {journal} {\bibinfo
  {journal} {Phys.Rev.}\ }\textbf {\bibinfo {volume} {D88}},\ \bibinfo {pages}
  {083009} (\bibinfo {year} {2013})},\ \Eprint {http://arxiv.org/abs/1305.0830}
  {arXiv:1305.0830 [astro-ph.HE]} \BibitemShut {NoStop}%
\bibitem [{\citenamefont {Cholis}\ \emph
  {et~al.}(2015{\natexlab{b}})\citenamefont {Cholis}, \citenamefont {Hooper},\
  and\ \citenamefont {Linden}}]{Cholis:2014lta}%
  \BibitemOpen
  \bibfield  {author} {\bibinfo {author} {\bibfnamefont {I.}~\bibnamefont
  {Cholis}}, \bibinfo {author} {\bibfnamefont {D.}~\bibnamefont {Hooper}}, \
  and\ \bibinfo {author} {\bibfnamefont {T.}~\bibnamefont {Linden}},\ }\href
  {\doibase 10.1088/1475-7516/2015/06/043} {\bibfield  {journal} {\bibinfo
  {journal} {JCAP}\ }\textbf {\bibinfo {volume} {1506}},\ \bibinfo {pages}
  {043} (\bibinfo {year} {2015}{\natexlab{b}})},\ \Eprint
  {http://arxiv.org/abs/1407.5625} {arXiv:1407.5625 [astro-ph.HE]} \BibitemShut
  {NoStop}%
\bibitem [{\citenamefont {Petrović}\ \emph {et~al.}(2015)\citenamefont
  {Petrović}, \citenamefont {Serpico},\ and\ \citenamefont
  {Zaharijas}}]{Petrovic:2014xra}%
  \BibitemOpen
  \bibfield  {author} {\bibinfo {author} {\bibfnamefont {J.}~\bibnamefont
  {Petrović}}, \bibinfo {author} {\bibfnamefont {P.~D.}\ \bibnamefont
  {Serpico}}, \ and\ \bibinfo {author} {\bibfnamefont {G.}~\bibnamefont
  {Zaharijas}},\ }\href {\doibase 10.1088/1475-7516/2015/02/023} {\bibfield
  {journal} {\bibinfo  {journal} {JCAP}\ }\textbf {\bibinfo {volume} {1502}},\
  \bibinfo {pages} {023} (\bibinfo {year} {2015})},\ \Eprint
  {http://arxiv.org/abs/1411.2980} {arXiv:1411.2980 [astro-ph.HE]} \BibitemShut
  {NoStop}%
\bibitem [{\citenamefont {Abdo}\ \emph {et~al.}(2013)\citenamefont {Abdo} \emph
  {et~al.}}]{TheFermi-LAT:2013ssa}%
  \BibitemOpen
  \bibfield  {author} {\bibinfo {author} {\bibfnamefont {A.}~\bibnamefont
  {Abdo}} \emph {et~al.} (\bibinfo {collaboration} {The Fermi-LAT
  collaboration}),\ }\href {\doibase 10.1088/0067-0049/208/2/17} {\bibfield
  {journal} {\bibinfo  {journal} {Astrophys.J.Suppl.}\ }\textbf {\bibinfo
  {volume} {208}},\ \bibinfo {pages} {17} (\bibinfo {year} {2013})},\ \Eprint
  {http://arxiv.org/abs/1305.4385} {arXiv:1305.4385 [astro-ph.HE]} \BibitemShut
  {NoStop}%
\bibitem [{\citenamefont {Lee}\ \emph {et~al.}(2015)\citenamefont {Lee},
  \citenamefont {Lisanti},\ and\ \citenamefont {Safdi}}]{Lee:2014mza}%
  \BibitemOpen
  \bibfield  {author} {\bibinfo {author} {\bibfnamefont {S.~K.}\ \bibnamefont
  {Lee}}, \bibinfo {author} {\bibfnamefont {M.}~\bibnamefont {Lisanti}}, \ and\
  \bibinfo {author} {\bibfnamefont {B.~R.}\ \bibnamefont {Safdi}},\ }\href
  {\doibase 10.1088/1475-7516/2015/05/056} {\bibfield  {journal} {\bibinfo
  {journal} {JCAP}\ }\textbf {\bibinfo {volume} {1505}},\ \bibinfo {pages}
  {056} (\bibinfo {year} {2015})},\ \Eprint {http://arxiv.org/abs/1412.6099}
  {arXiv:1412.6099 [astro-ph.CO]} \BibitemShut {NoStop}%
\bibitem [{\citenamefont {Cholis}\ \emph {et~al.}(2014)\citenamefont {Cholis},
  \citenamefont {Hooper},\ and\ \citenamefont {Linden}}]{Cholis:2014noa}%
  \BibitemOpen
  \bibfield  {author} {\bibinfo {author} {\bibfnamefont {I.}~\bibnamefont
  {Cholis}}, \bibinfo {author} {\bibfnamefont {D.}~\bibnamefont {Hooper}}, \
  and\ \bibinfo {author} {\bibfnamefont {T.}~\bibnamefont {Linden}},\
  }\href@noop {} {\  (\bibinfo {year} {2014})},\ \Eprint
  {http://arxiv.org/abs/1407.5583} {arXiv:1407.5583 [astro-ph.HE]} \BibitemShut
  {NoStop}%
\bibitem [{\citenamefont {Strong}(2007)}]{Strong:2006hf}%
  \BibitemOpen
  \bibfield  {author} {\bibinfo {author} {\bibfnamefont {A.~W.}\ \bibnamefont
  {Strong}},\ }\href {\doibase 10.1007/s10509-007-9480-1} {\bibfield  {journal}
  {\bibinfo  {journal} {Astrophys.Space Sci.}\ }\textbf {\bibinfo {volume}
  {309}},\ \bibinfo {pages} {35} (\bibinfo {year} {2007})},\ \Eprint
  {http://arxiv.org/abs/astro-ph/0609359} {arXiv:astro-ph/0609359 [astro-ph]}
  \BibitemShut {NoStop}%
\bibitem [{\citenamefont {Venter}\ \emph {et~al.}(2014)\citenamefont {Venter},
  \citenamefont {Johnson}, \citenamefont {Harding},\ and\ \citenamefont
  {Grove}}]{Venter:2014zea}%
  \BibitemOpen
  \bibfield  {author} {\bibinfo {author} {\bibfnamefont {C.}~\bibnamefont
  {Venter}}, \bibinfo {author} {\bibfnamefont {T.}~\bibnamefont {Johnson}},
  \bibinfo {author} {\bibfnamefont {A.}~\bibnamefont {Harding}}, \ and\
  \bibinfo {author} {\bibfnamefont {J.}~\bibnamefont {Grove}},\ }\href@noop {}
  {\  (\bibinfo {year} {2014})},\ \Eprint {http://arxiv.org/abs/1411.0559}
  {arXiv:1411.0559 [astro-ph.HE]} \BibitemShut {NoStop}%
\bibitem [{\citenamefont {Damiani}\ \emph {et~al.}(1997)\citenamefont
  {Damiani}, \citenamefont {Maggio}, \citenamefont {Micela},\ and\
  \citenamefont {Sciortino}}]{Damiani:1997}%
  \BibitemOpen
  \bibfield  {author} {\bibinfo {author} {\bibfnamefont {F.}~\bibnamefont
  {Damiani}}, \bibinfo {author} {\bibfnamefont {A.}~\bibnamefont {Maggio}},
  \bibinfo {author} {\bibfnamefont {G.}~\bibnamefont {Micela}}, \ and\ \bibinfo
  {author} {\bibfnamefont {S.}~\bibnamefont {Sciortino}},\ }\href
  {http://stacks.iop.org/0004-637X/483/i=1/a=350} {\bibfield  {journal}
  {\bibinfo  {journal} {The Astrophysical Journal}\ }\textbf {\bibinfo {volume}
  {483}},\ \bibinfo {pages} {350} (\bibinfo {year} {1997})}\BibitemShut
  {NoStop}%
\bibitem [{\citenamefont {Gonzalez-Nuevo}\ \emph {et~al.}(2006)\citenamefont
  {Gonzalez-Nuevo}, \citenamefont {Argueso}, \citenamefont {Lopez-Caniego},
  \citenamefont {Toffolatti}, \citenamefont {Sanz} \emph
  {et~al.}}]{GonzalezNuevo:2006pa}%
  \BibitemOpen
  \bibfield  {author} {\bibinfo {author} {\bibfnamefont {J.}~\bibnamefont
  {Gonzalez-Nuevo}}, \bibinfo {author} {\bibfnamefont {F.}~\bibnamefont
  {Argueso}}, \bibinfo {author} {\bibfnamefont {M.}~\bibnamefont
  {Lopez-Caniego}}, \bibinfo {author} {\bibfnamefont {L.}~\bibnamefont
  {Toffolatti}}, \bibinfo {author} {\bibfnamefont {J.}~\bibnamefont {Sanz}},
  \emph {et~al.},\ }\href {\doibase 10.1111/j.1365-2966.2006.10442.x}
  {\bibfield  {journal} {\bibinfo  {journal} {Mon.Not.Roy.Astron.Soc.}\
  }\textbf {\bibinfo {volume} {369}},\ \bibinfo {pages} {1603} (\bibinfo {year}
  {2006})},\ \Eprint {http://arxiv.org/abs/astro-ph/0604376}
  {arXiv:astro-ph/0604376 [astro-ph]} \BibitemShut {NoStop}%
\bibitem [{\citenamefont {Ade}\ \emph {et~al.}(2014)\citenamefont {Ade} \emph
  {et~al.}}]{Ade:2013jbr}%
  \BibitemOpen
  \bibfield  {author} {\bibinfo {author} {\bibfnamefont {P.}~\bibnamefont
  {Ade}} \emph {et~al.} (\bibinfo {collaboration} {Planck}),\ }\href {\doibase
  10.1051/0004-6361/201321524} {\bibfield  {journal} {\bibinfo  {journal}
  {Astron.Astrophys.}\ }\textbf {\bibinfo {volume} {571}},\ \bibinfo {pages}
  {A28} (\bibinfo {year} {2014})},\ \Eprint {http://arxiv.org/abs/1303.5088}
  {arXiv:1303.5088 [astro-ph.CO]} \BibitemShut {NoStop}%
\bibitem [{\citenamefont {GrŽgoire}\ and\ \citenamefont
  {Knšdlseder}(2013)}]{Gregoire:2013yta}%
  \BibitemOpen
  \bibfield  {author} {\bibinfo {author} {\bibfnamefont {T.}~\bibnamefont
  {GrŽgoire}}\ and\ \bibinfo {author} {\bibfnamefont {J.}~\bibnamefont
  {Knšdlseder}},\ }\href {\doibase 10.1051/0004-6361/201219676} {\bibfield
  {journal} {\bibinfo  {journal} {Astron. Astrophys.}\ }\textbf {\bibinfo
  {volume} {554}},\ \bibinfo {pages} {A62} (\bibinfo {year} {2013})},\ \Eprint
  {http://arxiv.org/abs/1305.1584} {arXiv:1305.1584 [astro-ph.GA]} \BibitemShut
  {NoStop}%
\bibitem [{\citenamefont {Calore}\ \emph {et~al.}(2014)\citenamefont {Calore},
  \citenamefont {Di~Mauro}, \citenamefont {Donato},\ and\ \citenamefont
  {Donato}}]{Calore:2014oga}%
  \BibitemOpen
  \bibfield  {author} {\bibinfo {author} {\bibfnamefont {F.}~\bibnamefont
  {Calore}}, \bibinfo {author} {\bibfnamefont {M.}~\bibnamefont {Di~Mauro}},
  \bibinfo {author} {\bibfnamefont {F.}~\bibnamefont {Donato}}, \ and\ \bibinfo
  {author} {\bibfnamefont {F.}~\bibnamefont {Donato}},\ }\href {\doibase
  10.1088/0004-637X/796/1/14} {\bibfield  {journal} {\bibinfo  {journal}
  {Astrophys.J.}\ }\textbf {\bibinfo {volume} {796}},\ \bibinfo {pages} {1}
  (\bibinfo {year} {2014})},\ \Eprint {http://arxiv.org/abs/1406.2706}
  {arXiv:1406.2706 [astro-ph.HE]} \BibitemShut {NoStop}%
\bibitem [{\citenamefont {Topchiev}\ \emph {et~al.}(2015)\citenamefont
  {Topchiev}, \citenamefont {Galper}, \citenamefont {Bonvicini}, \citenamefont
  {Adriani}, \citenamefont {Aptekar} \emph {et~al.}}]{gamma400}%
  \BibitemOpen
  \bibfield  {author} {\bibinfo {author} {\bibfnamefont {N.}~\bibnamefont
  {Topchiev}}, \bibinfo {author} {\bibfnamefont {A.}~\bibnamefont {Galper}},
  \bibinfo {author} {\bibfnamefont {V.}~\bibnamefont {Bonvicini}}, \bibinfo
  {author} {\bibfnamefont {O.}~\bibnamefont {Adriani}}, \bibinfo {author}
  {\bibfnamefont {R.}~\bibnamefont {Aptekar}},  \emph {et~al.},\ }\href
  {\doibase 10.3103/S1062873815030429} {\bibfield  {journal} {\bibinfo
  {journal} {Bull.Russ.Acad.Sci.Phys.}\ }\textbf {\bibinfo {volume} {79}},\
  \bibinfo {pages} {417} (\bibinfo {year} {2015})}\BibitemShut {NoStop}%
\bibitem [{\citenamefont {Wu}\ \emph {et~al.}(2014)\citenamefont {Wu},
  \citenamefont {Su}, \citenamefont {Bravar}, \citenamefont {Chang},
  \citenamefont {Fan} \emph {et~al.}}]{pangu}%
  \BibitemOpen
  \bibfield  {author} {\bibinfo {author} {\bibfnamefont {X.}~\bibnamefont
  {Wu}}, \bibinfo {author} {\bibfnamefont {M.}~\bibnamefont {Su}}, \bibinfo
  {author} {\bibfnamefont {A.}~\bibnamefont {Bravar}}, \bibinfo {author}
  {\bibfnamefont {J.}~\bibnamefont {Chang}}, \bibinfo {author} {\bibfnamefont
  {Y.}~\bibnamefont {Fan}},  \emph {et~al.},\ }\href {\doibase
  10.1117/12.2057251} {\bibfield  {journal} {\bibinfo  {journal} {Proc.SPIE
  Int.Soc.Opt.Eng.}\ }\textbf {\bibinfo {volume} {9144}},\ \bibinfo {pages}
  {91440F} (\bibinfo {year} {2014})},\ \Eprint {http://arxiv.org/abs/1407.0710}
  {arXiv:1407.0710 [astro-ph.IM]} \BibitemShut {NoStop}%
\bibitem [{\citenamefont {Lee}\ \emph {et~al.}(2016)\citenamefont {Lee},
  \citenamefont {Lisanti}, \citenamefont {Safdi}, \citenamefont {Slatyer},\
  and\ \citenamefont {Xue}}]{Lee:2015fea}%
  \BibitemOpen
  \bibfield  {author} {\bibinfo {author} {\bibfnamefont {S.~K.}\ \bibnamefont
  {Lee}}, \bibinfo {author} {\bibfnamefont {M.}~\bibnamefont {Lisanti}},
  \bibinfo {author} {\bibfnamefont {B.~R.}\ \bibnamefont {Safdi}}, \bibinfo
  {author} {\bibfnamefont {T.~R.}\ \bibnamefont {Slatyer}}, \ and\ \bibinfo
  {author} {\bibfnamefont {W.}~\bibnamefont {Xue}},\ }\href {\doibase
  10.1103/PhysRevLett.116.051103} {\bibfield  {journal} {\bibinfo  {journal}
  {Phys. Rev. Lett.}\ }\textbf {\bibinfo {volume} {116}},\ \bibinfo {pages}
  {051103} (\bibinfo {year} {2016})}\BibitemShut {NoStop}%
\bibitem [{\citenamefont {Dame}\ \emph {et~al.}(2001)\citenamefont {Dame},
  \citenamefont {Hartmann},\ and\ \citenamefont {Thaddeus}}]{Dame:2000sp}%
  \BibitemOpen
  \bibfield  {author} {\bibinfo {author} {\bibfnamefont {T.}~\bibnamefont
  {Dame}}, \bibinfo {author} {\bibfnamefont {D.}~\bibnamefont {Hartmann}}, \
  and\ \bibinfo {author} {\bibfnamefont {P.}~\bibnamefont {Thaddeus}},\ }\href
  {\doibase 10.1086/318388} {\bibfield  {journal} {\bibinfo  {journal}
  {Astrophys.J.}\ }\textbf {\bibinfo {volume} {547}},\ \bibinfo {pages} {792}
  (\bibinfo {year} {2001})},\ \Eprint {http://arxiv.org/abs/astro-ph/0009217}
  {arXiv:astro-ph/0009217 [astro-ph]} \BibitemShut {NoStop}%
\bibitem [{\citenamefont {Casandjian}(2015)}]{Casandjian:2015hja}%
  \BibitemOpen
  \bibfield  {author} {\bibinfo {author} {\bibfnamefont {J.-M.}\ \bibnamefont
  {Casandjian}},\ }\href {\doibase 10.1088/0004-637X/806/2/240} {\bibfield
  {journal} {\bibinfo  {journal} {Astrophys. J.}\ }\textbf {\bibinfo {volume}
  {806}},\ \bibinfo {pages} {240} (\bibinfo {year} {2015})},\ \Eprint
  {http://arxiv.org/abs/1506.00047} {arXiv:1506.00047 [astro-ph.HE]}
  \BibitemShut {NoStop}%
\bibitem [{Note1()}]{Note1}%
  \BibitemOpen
  \bibinfo {note} {We note for reference that for such a MSP, placed at GC
  distance, we would have seen around 270 photons in our energy range. From
  this and Tab.~\ref {tab:sources} one can estimate that $\sim 100$ photons
  correspond to a wavelet signal with a significance of $\protect \mathcal
  {S}\sim 2$.}\BibitemShut {Stop}%
\bibitem [{\citenamefont {Stark}\ and\ \citenamefont
  {Lee}(2005)}]{Stark:2004bs}%
  \BibitemOpen
  \bibfield  {author} {\bibinfo {author} {\bibfnamefont {A.~A.}\ \bibnamefont
  {Stark}}\ and\ \bibinfo {author} {\bibfnamefont {Y.-G.}\ \bibnamefont
  {Lee}},\ }\href {\doibase 10.1086/427936} {\bibfield  {journal} {\bibinfo
  {journal} {Astrophys. J.}\ }\textbf {\bibinfo {volume} {619}},\ \bibinfo
  {pages} {L159} (\bibinfo {year} {2005})},\ \Eprint
  {http://arxiv.org/abs/astro-ph/0403631} {arXiv:astro-ph/0403631 [astro-ph]}
  \BibitemShut {NoStop}%
\end{thebibliography}%

\clearpage
\onecolumngrid
\section{Supplemental Material}
In the Supplemental Material we discuss the possible impact of various
systematic effects on our results.  This includes a control region analysis, a
discussion of various types of \gr\ sources, substructure in diffuse
emission, a thick-disk population of MSPs, sphericity and the role of negative
wavelet peaks.

\subsection{A. Null results in control regions}
\renewcommand{\thefigure}{S-\arabic{figure}}
\setcounter{figure}{0}

In order to estimate the effect of various systematic uncertainties, it is
useful to apply our analysis on control regions along the Galactic disk (in the
case of Galactic diffuse emission this was first systematically done in
Ref.~\cite{Calore:2014xka}).  Potentially unresolved substructure in the
Galactic diffuse emission (e.g.~in the form of giant molecular clouds, see
below), and contributions from various Galactic and extragalactic source
populations could be responsible for the detected wavelet signal in the inner
Galaxy, but would in general also affect other regions in the Galactic disk.
To this end, we focus on (partially overlapping) control regions along the
Galactic disk, which are of the same size as the inner Galaxy ROI, but
displaced by $\Delta \ell = \pm k\, 20^\circ$ with $k=1,2,3,4$.

\begin{figure}[h]
  \begin{center}
        \includegraphics[width=0.55\linewidth]{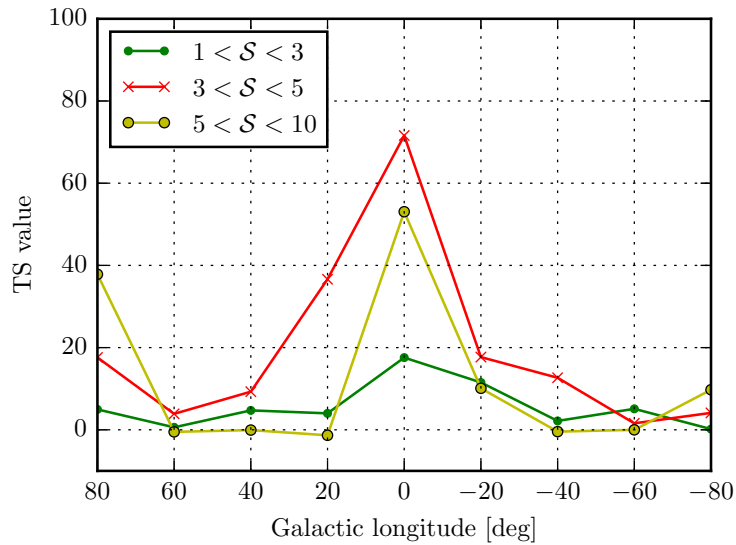}
  \end{center}
  \caption{We show, as function of the central longitudinal position of the
    control regions (for $\ell=0^\circ$ the main ROI), the significance of a
    CSP detection.  We assume that the CSP is centered in each of
    the ROIs, and we refit the parameters $\Lmax$ and the number of sources.
    We indicate how different ranges of the wavelet peak significances
    contribute to the detection.  \emph{All} 3FGL sources are masked in this
    plot, in order to be conservative.}
  \label{fig:TSs}
\end{figure}

In Fig.~\ref{fig:TSs} we show the $TS$ value for a detection of the CSP for the
main and the different control ROIs along the Galactic disk.  We leave $\Lmax$
and $\Phi_5$ free to vary in each region independently.  In the main ROI that
covers the inner Galaxy we find the significant detection of a CSP that was
discussed in the main text.  As shown in the plot, this high significance is
supported by the low, intermediate and high-significance SNR peaks of the
wavelet transform separately.  The directly adjacent regions also show
relatively large $TS$ values, which is either caused by the partial overlap of
these control regions with the main ROI, or by a CSP that is more disk-like
than assumed in our analysis.  We will address the latter point below.
However, in the outermost six control regions we find no significant detection
of a CSP, for any of the considered values for $\Lmax$ and $\Phi_5$ (the large
$TS$ values at $\ell=80^\circ$ are caused by one extremely bright source that
generates fake peaks in its tails).  This observation makes it already
extremely unlikely that our findings are driven by a mismodelling of the local
Galactic diffuse emission, or by extragalactic sources.  We will address this
in more detail below.

\subsection{B. Consistent wavelet signal in separate bins}

\begin{figure}
  \begin{center}
        \includegraphics[width=0.55\linewidth]{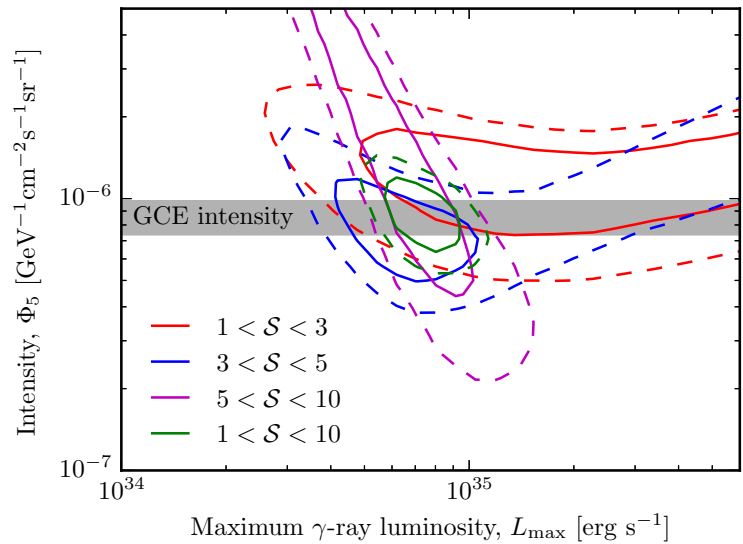}
  \end{center}
  \caption{Similar to Fig.~3 in the main text, but showing limits derived
    for different ranges of the wavelet peak significances separately.  The
    parameter degeneracies present in some of the fits cancel when fitting the
    entire significance range at once.  We only show $68.7\%$ and $95.4\%$ CL
    constraints for clarity.}
  \label{fig:limitsBG}
\end{figure}

It is instructive to see how wavelet peaks with different significances
contribute to the constraints on the luminosity function that we showed in
Fig.~\ref{fig:limits} in the main text. To this end, we show in
Fig.~\ref{fig:limitsBG} the limits that we obtain separately from peak
significances in the range $\mathcal{S}=1$--$3$, $\mathcal{S}=3$--$5$ and
$\mathcal{S}=5$--$10$, respectively.  All three constraints are mutually
consistent to within $1\sigma$, leading to a consistent interpretation of the
peaks shown in Fig.~\ref{fig:transform} in the main text.  In all cases we find
some degeneracy in the ($\Lmax$, $\Phi_5$) plane.  High significance peaks
predominantly provide a stringent upper limit on $\Lmax$, whereas low
significance peaks mostly constrain the overall luminosity of the modelled
source population.  

\begin{figure}
    \begin{center}
      \includegraphics[width=0.55\linewidth]{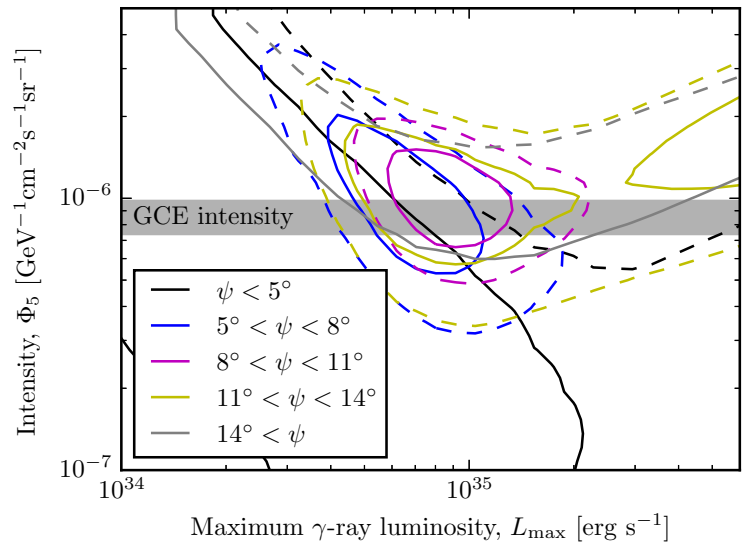}
    \end{center}
    \caption{Similar to Fig.~3 in the main text, but showing
    limits as derived for different spatial bins separately.}
    \label{fig:limits_rings}
\end{figure}

To show that our assumption on the spatial distribution of the CSP is
consistent with the data, we show Fig.~\ref{fig:limits_rings} the result
obtained for the five different spatial bins independently.  Ring 1--5
correspond to $r \in \left[i^\circ, i+3^\circ\right]$ with $i = 2, 5, 8, 11,
14$, respectively.  We find constraints on $\Lmax$ and $\Phi_5$ that are
mostly consistent to within 1$\sigma$.

Finally, we checked that the identified wavelet peaks are symmetrically
distributed in the north, south, east and west parts of our main ROI.  Only at
$S>3$ we find a slight (statistically not very significant) asymmetry with more
peaks in the south, which might be caused by the somewhat stronger Galactic
foregrounds in the north, which makes point source detection in the north more
challenging.

\subsection{C. Mild dependence on the MSP luminosity function}

\begin{figure}
  \begin{center}
        \includegraphics[width=0.55\linewidth]{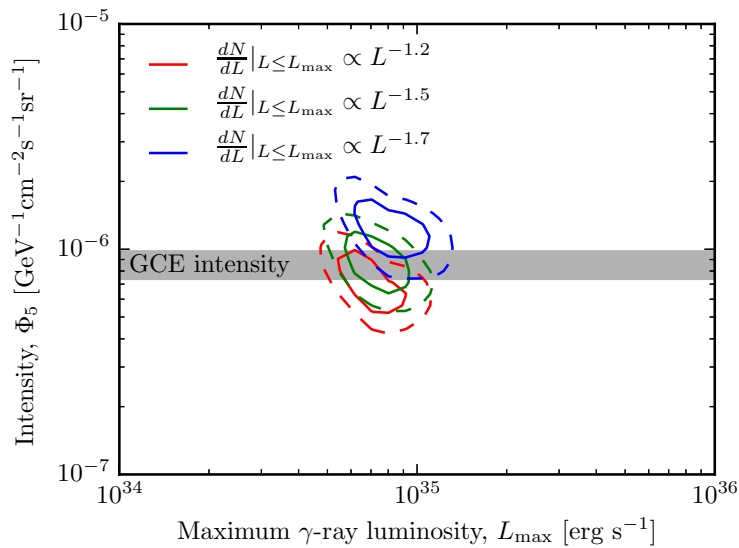}
  \end{center}
  \caption{Similar to Fig.~3 in the main text. We show the
  $68.7\%$ and $95.4\%$ CL contours for luminosity functions with spectral
  indices of $1.2$, $1.5$ and $1.7$.}
  \label{fig:limitsAlpha}
\end{figure}

Theoretically, $\alpha$ is not well constrained and can plausibly range from
$1.5$ to $3$, depending on the emission model~\cite{Strong:2006hf,
Venter:2014zea, Petrovic:2014xra}.  Actual MSP observations actually seem to
indicate somewhat smaller values closer to $\alpha\sim
1.2$~\cite{Cholis:2014noa}.  We show in Fig.~\ref{fig:limitsAlpha} the $68.7\%$
and $95.4\%$ CL contours for different luminosity functions, respectively with
spectral indices of $1.2$ and $1.7$. For a fixed intensity ($\Phi_5$),
hardening (softening) of the luminosity function corresponds to an enhancement
(suppression) of the number of sub-threshold point sources, which explains the
direction in which the best fit region moves.  We note that we obtain very
similar $TS$ values for all slopes that we considered.

\subsection{D. The role of unmasked 3FGL sources}

\begin{table}[t]
    \begin{tabular}{lrccccccr}
        \hline
        \hline
        3FGL Name && $\ell\,[^\circ]$ & $b\,[^\circ]$ && $\chi^2/\rm dof$ &
        $\sqrt{TS}$ & $\mathcal{S}$ & $L\, [10^{34}\,\rm erg/s]$\\[1pt]
        \hline
        J1649.6-3007 && -7.99 & 9.27 && 1.07 & 5.6 & 7.4 & $7.8^{+5.5}_{-2.7}$ \\
        J1703.6-2850 && -5.08 & 7.65 && 0.48 & 2.4 & 5.0 & $3.4^{+2.4}_{-1.2}$ \\
        J1740.5-2642 && 1.30 & 2.12 && 0.37 & 6.4 & 3.5 & $14.9^{+10.6}_{-5.1}$ \\
        J1740.8-1933 && 7.43 & 5.83 && 0.77 & 1.9 & 2.3 & $3.8^{+2.7}_{-1.3}$ \\
        J1744.8-1557 && 11.03 & 6.88 && 0.40 & 3.7 & 3.4 & $5.6^{+3.9}_{-1.9}$ \\
        J1758.8-4108 && -9.21 & -8.48 && 0.90 & 5.6 & 3.8 & $4.8^{+3.4}_{-1.7}$ \\
        J1759.2-3848 && -7.11 & -7.43 && 0.35 & 4.6 & 5.6 & $5.9^{+4.2}_{-2.0}$ \\
        J1808.3-3357 && -1.94 & -6.71 && 0.40 & 6.9 & 6.3 & $8.0^{+5.7}_{-2.7}$ \\
        J1808.4-3519 && -3.15 & -7.36 && 0.41 & 4.6 & 4.4 & $5.0^{+3.5}_{-1.7}$ \\
        J1808.4-3703 && -4.68 & -8.19 && 0.22 & 4.9 & 5.3 & $4.3^{+3.1}_{-1.5}$ \\
        J1820.4-3217 && 0.74 & -8.17 && 1.04 & 5.7 & 1.7 & $7.2^{+5.1}_{-2.5}$ \\
        J1830.8-3136 && 2.35 & -9.84 && 0.54 & 5.9 & 6.0 & $5.0^{+3.6}_{-1.7}$ \\
        J1837.3-2403 && 9.85 & -7.81 && 0.28 & 4.0 & 3.0 & $4.7^{+3.3}_{-1.6}$ \\
        \hline
        \hline
    \end{tabular}
    \caption{List of the 13 unassociated 3FGL sources with MSP-like spectra,
      which we leave unmasked in our analysis.  If the GeV excess is caused by
      dim point sources, it is likely that some or most of
      them are part of the CSP.  The last four columns show the goodness-of-fit
      of the reference MSP spectrum, the 3FGL significance in the 1--3 GeV
      band, the corresponding peak of the wavelet SNR, and the \gr\ luminosity
      (assuming $8.5\pm2$ kpc distance from the source and the reference stacked
      MSP spectrum from the main paper with a normalization that is obtained
      from a fit to the measured source flux).}
    \label{tab:sources}
\end{table}

In the present analysis, we make use of the 3FGL, the third \Fermi\ source
catalogue, which is based on the first four years of \Fermi\ pass 7 data.  One
important ingredient in our analysis is the masking of 3FGL sources.  These
sources are of Galactic and extragalactic origin and leaving them unmasked
would inevitably induce sizeable signals in our search for a sub-threshold
source population in the bulge.  However, as discussed in the main text, we
keep unassociated sources with MSP-like spectra unmasked.  These sources could
be part of the bulge MSP population that we are looking for, and masking them
would bias our results.  The 13 sources that pass our MSP cuts are listed in
Tab.~\ref{tab:sources}.  

\begin{figure}
  \begin{center}
    \includegraphics[width=0.55\linewidth]{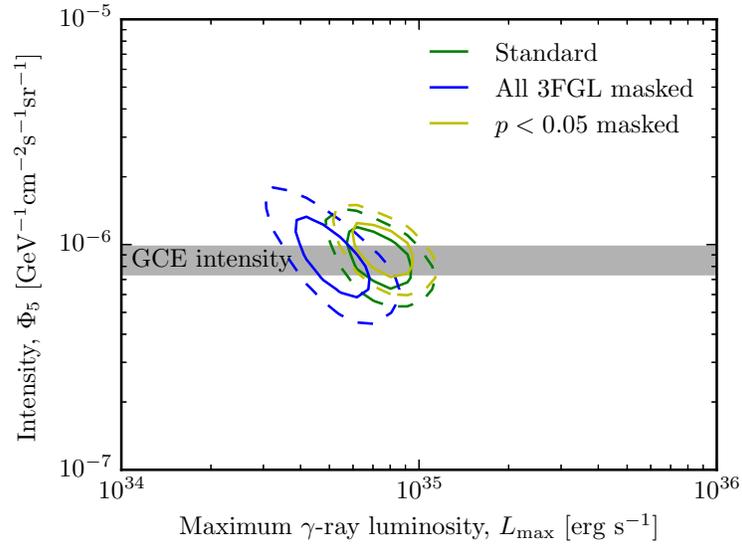}
  \end{center}
  \caption{Similar to Fig.~3 in the main text. We show the effect of masking
  \emph{all} 3FGL sources, or leaving a slightly larger number unmasked.}
  \label{fig:limitsPvalues}
\end{figure}

In general, falsely masking unassociated sources that actually belong to the
bulge MSP population would push $\Lmax$ to lower values, whereas falsely
unmasking foreground sources would push it to large values.  This is
illustrated in Fig.~\ref{fig:limitsPvalues}.  We show the case where we mask
\emph{all} unassociated sources, as well as the case where we adopt a weaker
criterion for the spectral fit, leaving around 20 sources unmasked.  We find
that in both cases the best-fit value for $\Lmax$ moves in the expected
direction, but the results remain consistent to within $1\sigma$.  Furthermore,
the significance of our wavelet detection that we quote in the main text (where
we include $\mathcal{S}<5$ bins only) changes to $9.2\sigma$ when we mask all
sources, and to $10.8\sigma$ when keeping 20 sources unmasked.  The
(un-)masking of 3FGL is hence not decisive for our qualitative findings,
although quantitative results can be affected.

\medskip

It is interesting to note that the faintest 3FGL source in the inner Galaxy ROI
that passes our MSP-spectrum cut, has a luminosity of $L=3.4\times10^{34}\ergs$
if placed at 8.5 kpc distance.  This is a good, though rough, indication for
the \emph{de facto} sensitivity threshold of \Fermi-LAT for the detection of
sources with MSP-like spectra in the bulge region.

Lastly, one can use the sources in Tab.~\ref{tab:sources} to compare the
sensitivity of the 3FGL with our wavelet analysis.  Averaging over the 13
sources, we find a ratio of $\mathcal{S}/\sqrt{TS}\simeq 1.0\pm0.4$, indicating
that the sensitivity of the wavelet method is similar to the 3FGL sensitivity
in a comparable energy range.  However, the scatter exceeds the one expected
from statistical fluctuations alone, which can be attributed to differences in
the systematics that affect the 3FGL and the wavelet analysis.

\subsection{E. The role of various source populations}

\begin{figure}
  \begin{center}
        \includegraphics[width=0.55\linewidth]{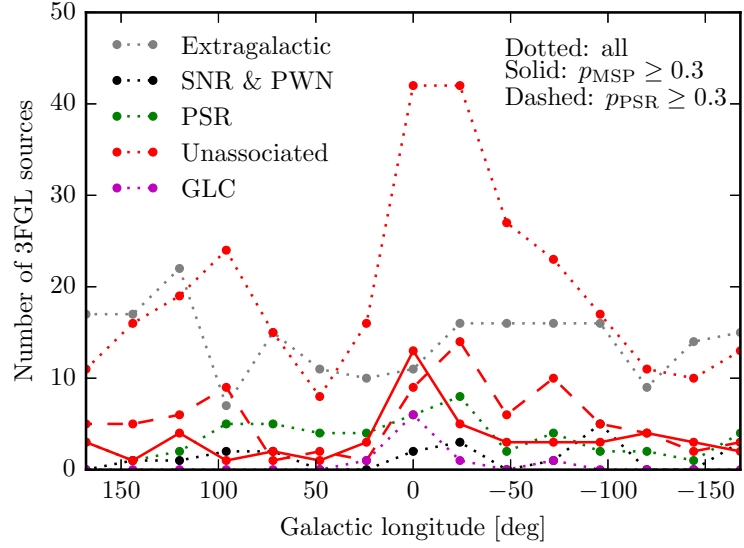}
  \end{center}
  \caption{Latitude profile of sources, for different ROIs along the Galactic
    disk, as function of their longitudinal center (dotted lines).  The ROIs have a size of
    $24^\circ\times24^\circ$, with the Galactic disk, $|b|<2^\circ$, masked.
    The different colors correspond to different source categories.
    In the case of unassociated sources, the solid (dashed) lines show only sources that pass the spectrum
    cut for MSP-like (young pulsar-like) sources.}
  \label{fig:sources}
\end{figure}

\begin{figure}
  \begin{center}
    \includegraphics[width=0.55\linewidth]{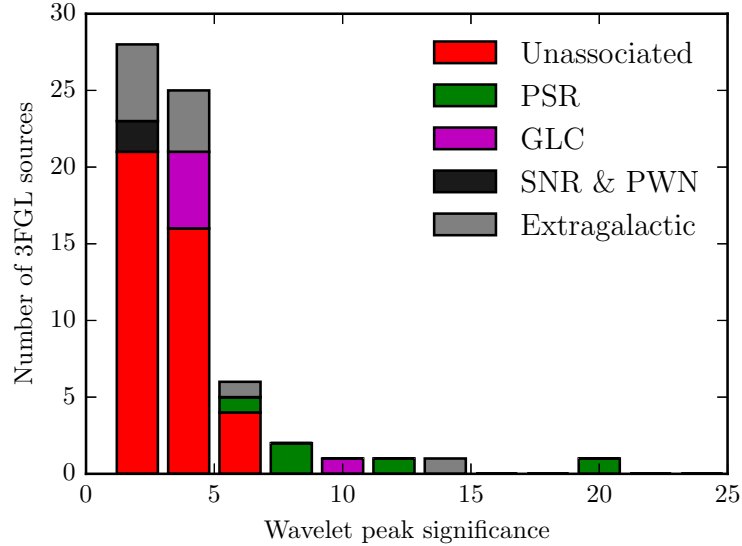}
  \end{center}
  \caption{Stacked histogram of wavelet peak values $\mathcal{S}$ that correspond to
    3FGL sources in the inner Galaxy ROI. We show the contribution from
    different source categories separately.  Unassociated sources generate
    mostly low-significance wavelet peaks, whereas for example source that are
    marked as pulsar in the 3FGL only generate
    peaks with a high significance (see discussion).}
  \label{fig:sources_hist}
\end{figure}

In Fig.~\ref{fig:sources}, we show the number of 3FGL sources in our inner
Galaxy ROI as well as in various same-sized ROIs that are displaced along the
Galactic disk in steps of $\Delta\ell=\pm 24^\circ$.  We show (identified and
associated) extragalactic sources, various Galactic source classes and
unassociated sources classes separately.  We also show for the unassociated sourced
how many sources pass our MSP cut.  It is apparent that the number
of unassociated sources strongly peaks in the inner Galaxy ROI, however with a
clear asymmetry towards negative values of $\ell$, and another peak around
$\ell\approx 100^\circ$.  However, after applying the MSP cut, predominantly
sources in the inner Galaxy survive.  In Fig.~\ref{fig:sources_hist}, on the
other hand, we show a histogram of the wavelet peak significance that our
analysis attributes to the 3FGL sources in the inner Galaxy ROI.  Again,
unassociated sources play a major role and produce wavelet peaks down to values
of $\mathcal{S}\sim 1$.  On the other hand, 3FGL sources that are identified as
pulsars appear only with $\mathcal{S}>5$ in our analysis.  We will discuss in
the following the potential impact of each of the source classes separately.

\medskip

\paragraph{Extragalactic sources.} As shown in Fig.~\ref{fig:sources}, the
number of extragalactic sources in the individual ROIs along the Galactic disk
fluctuates around values of about $\sim13$.  No significant suppression is
observed in the inner Galaxy ROI, which would have indicated that it is more
challenging to identify or associate extragalactic sources in this region.  We
will use here a simple argument to show that extragalactic sources cannot play
a significant role for our results.  The average number of $\mathcal{S}=3$--$5$
wavelet peaks in the above control regions along the Galactic disk is 20; in
the inner Galaxy ROI it is 42  (we exclude here \emph{all} 3FGL sources to be
conservative).  If extragalactic sources were the main contribution to these
peaks, in addition to the about 10 wavelet peaks that are expected from
statistical fluctuations alone (see Fig.~\ref{fig:hist} in the main text), the
42 observed peaks would constitute a $>5\sigma$ upward fluctuation above the
expected 20.  This makes it extremely unlikely that extragalactic sources
contribute significantly to our results in the inner Galaxy.  Similar arguments
can be made using wavelet peaks in the range $\mathcal{S}=1$--$2$.  

\paragraph{Supernova remnants and pulsar wind nebulae.}  As apparent in
Fig.~\ref{fig:sources}, only a very small number of sources along the Galactic
disk at latitudes $|b|>2^\circ$ (and almost none at $|b|>5^\circ$) are
identified with supernova remnants or pulsar wind nebulae.  In our ROIs their
number is much less than the number of extragalactic sources, and their
distribution is centrally not peaked, indicating that sources at these
latitudes are mostly local.  Sources in this category are typically more easily
detectable at higher and lower energies than the energy range used in our
analysis, and would be most likely listed in the 3FGL and hence masked if they
were abundant and significant.  We consider it hence as extremely unlikely that
sources of this category significantly affect our results in the inner Galaxy.

\paragraph{Young and millisecond pulsars.}  An interesting feature of pulsars
in the 3FGL is that they always induce large wavelet signals in our analysis,
as shown in Fig.~\ref{fig:sources_hist}.  This makes indeed sense, since the
identification of a \gr\ source as pulsar requires the measurement of its
pulsation, and hence a large enough number of photons.  Furthermore, the pulsar
energy spectrum often peaks close the energy range of our analysis.  Already
from Fig.~\ref{fig:sources_hist} it is obvious that a large fraction of the
unassociated sources, which mostly appear with lower significance peaks, must
in fact be pulsars.  Interestingly, as shown in Fig.~\ref{fig:sources},
identified pulsars do not show a centrally peaked distribution, whereas the
unassociated sources clearly do.  This behaviour is expected, given that with
increasing distance the pulsar identification becomes more challenging.

\paragraph{Globular clusters.} The \gr\ emission from globular clusters that is
not masked in our analysis, either because the globular clusters did not enter
the 3FGL, or because they happen to be among our 13 unmasked unassociated
sources, could in principle contribute to the detected signal.  Since their
total emission is usually due to several MSPs which appear as a single source
for \Fermi-LAT, their presence could bias $\Lmax$ towards larger values.
However, given the simulation results from Ref.~\cite{Brandt:2015ula}, we
expect this effect to be small, and leave a more detailed discussion to future
work.

\paragraph{Unassociated sources.} The peak of unassociated sources in the inner
Galaxy, as shown in Fig.~\ref{fig:sources}, appears clearly asymmetric, with a
second peak at $\ell\approx 100^\circ$.  As discussed above, a large fraction
of these sources is expected to be young or millisecond pulsars.  Obviously,
the 3FGL unassociated sources do not directly contribute to our results since
they are masked (except the 13 MSP candidates).  However, the unassociated
sources are extremely abundant even down to $\mathcal{S}\sim1$, and their
probable nature can be used as an indicator for what source population
dominates just below threshold.

In Fig.~\ref{fig:sources}, we see that our MSP cut removes most of the
unassociated sources, but leaves an excess of 13 unassociated sources in the
inner Galaxy, as discussed above.  What is more, if we slightly modify the
spectral criterion, using $dN/dE\propto e^{-E/2\rm\, GeV} E^{-2}$, which is
somewhat more pulsar-like (softer index, lower cutoff), this behaviour changes
and we instead find excesses that are more correlated with the peaks of the
unassociated sources away from the inner Galaxy.  Although the statistical
significance of this finding is rather difficult to quantify without a detailed
study (which we leave for future work), this result is indicative.  It suggests
that a large fraction of the inner Galaxy unassociated (and sub-threshold)
sources are likely MSPs, whereas unassociated sources in other parts of the
disk have a larger fraction of young pulsars.  The latter point is further
supported by the fact that similar structures can be found in the longitudinal
distribution of identified pulsars.

In summary, we expect that our wavelet signal is dominated by whatever source
class is responsible for most of the unassociated sources towards the inner
Galaxy.  Very likely, these are millisecond and young pulsars, with a somewhat
higher MSP/young pulsar ratio than in the rest of the disk.  Since these
sources appear in general both in the Galactic disk as well as in the bulge, it
is important to study whether the excess/suppression of wavelet peaks in the
inner Galaxy points to a disk population, a bulge population, or to a
combination of both.  This will be discussed in the section G below.

\subsection{F. Possible caveats concerning the Galactic diffuse emission}

\begin{figure}
    \begin{center}
        \includegraphics[width=0.55\linewidth]{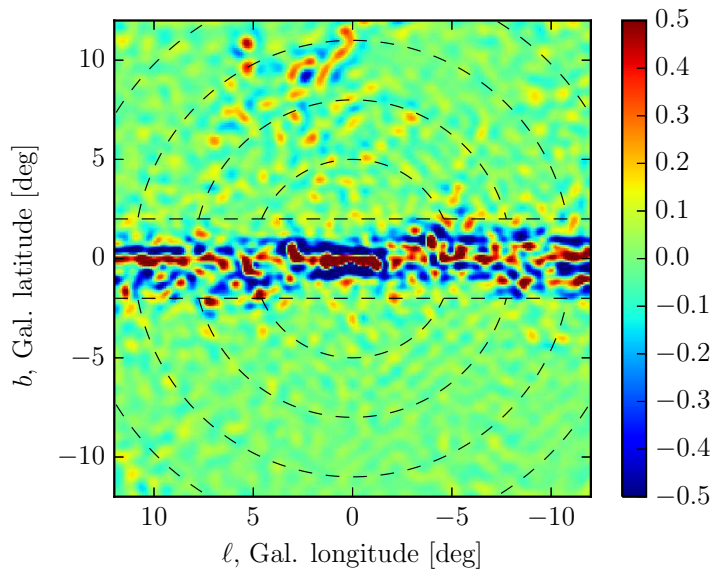}
    \end{center}
    \caption{Similar to Fig.~1 in the main text (note the
      different color scale), but showing the transform of the diffuse BG model
      only, without Poisson noise being applied.  Outside of the Galactic disk,
      $|b|>2^\circ$, which we exclude from our analysis, the significance of
      wavelet peaks remains below $0.5$. The variance is below $0.1$, which shows
      that even $1\sigma$ peaks in the wavelet transform are unlikely to be
      strongly affected by the Galactic diffuse emission.}
    \label{fig:transformWOpoisson}
\end{figure}

In our Monte Carlo studies, we use the standard \Fermi\ diffuse model for pass
8 data analysis.  The wavelet transform of this model, \emph{without applying
Poisson noise}, but using the same exposure as in our main analysis,  is shown
in Fig.~\ref{fig:transformWOpoisson}.  Outside of the masked Galactic disk at
$|b|>2^\circ$, we do not find any excesses with a significance larger than
about $0.5\sigma$.  The main effect of such variations would be to offset the
significance of random statistical fluctuations and sub-threshold point sources
towards higher or lower values.  This would not significantly affect peaks with
a large SNR.  However, it can potentially be important for low-significance
peaks, because the collective shift of a large number of peaks, even by a small
amount, could become statistically relevant.  But since the variance of
the SNR values shown in Fig.~\ref{fig:transformWOpoisson} is below 0.1, we do
not expect that the details of the modeling of diffuse emission when doing MCs
is going to affect our results.  

The \Fermi\ diffuse models might not actually contain all relevant small-scale
gas structures, and the effect of these missing structures on our results is not
straightforward to estimate without a detailed analysis and modeling of the
power-spectrum of gas at small scales.  It is hence rather important that our
non-detection of strong wavelet signals along the Galactic disk in
Fig.~\ref{fig:TSs} largely excludes that mismodeling of \emph{local} gas is the
cause for the detected signal towards the inner Galaxy, since it would affect
other parts of the disk as well.  This is in particular true since there is
relatively little molecular gas in our main ROI, compared to the control
regions~\cite{Dame:2000sp}.  Thus, gas-related effects should be larger in the
control regions than in the main ROI.  

If one insists on a gas-related interpretation, our results hence suggest that
the wavelet signals are caused by unmodeled gas in the Galactic bulge, at a
height of 0.3--$1.5\kpc$.  If we assume a cosmic-ray density in the bulge similar
to the local one, the differential \gr\ emissivity at 1 GeV is around
$3\times 10^{-26}\ \rm s^{-1} GeV^{-1}$ per hydrogen
atom~\cite{Casandjian:2015hja}.  This implies that dense gas clouds with masses
around $3\times10^5 {\rm\, M_\odot}$ would be at 1 GeV roughly as bright as
MSPs with a luminosity of $L = 7\times10^{34}\ergs$.\footnote{We note for
reference that for such a MSP, placed at GC distance, we would have seen
around 270 photons in our energy range.  From this and Tab.~\ref{tab:sources}
one can estimate that $\sim100$ photons correspond to a wavelet signal with a
significance of $\mathcal{S}\sim2$.} Interestingly, giant molecular clouds are
known objects of that mass, and they can be dense enough to appear at GC
distance point-like for \Fermi.  However, the scale height of known giant
molecular clouds is at the level of a few 10~pc (they usually intersect with
the Galactic disk) instead of the required $\sim1\kpc$~\cite{Stark:2004bs}.
Furthermore, clouds of that size should give rise to CO emission in the range
$\mathcal{O}(10$--$100)\ \rm K\, km\, s^{-1}$, which is not seen in current
observations~\cite{Dame:2000sp}.  If observed, such CO emission should be
distributed north/south symmetric, as our wavelet peaks are too.

If one could show that a large number of such giant molecular clouds (or other
structures with similar mass and density) can form and be transported to kpc
heights in the Galactic bulge, while hiding from all observations, the
interpretation of the identified wavelet peaks in terms of unmodeled gas would
remain a possibility.  However, as of now, and for all of the above reasons, we
regard gas-related interpretations of our results as rather unlikely and
speculative.

\subsection{G. Potential impact of thick-disk population}

\begin{figure}
  \begin{center}
        \includegraphics[width=0.55\linewidth]{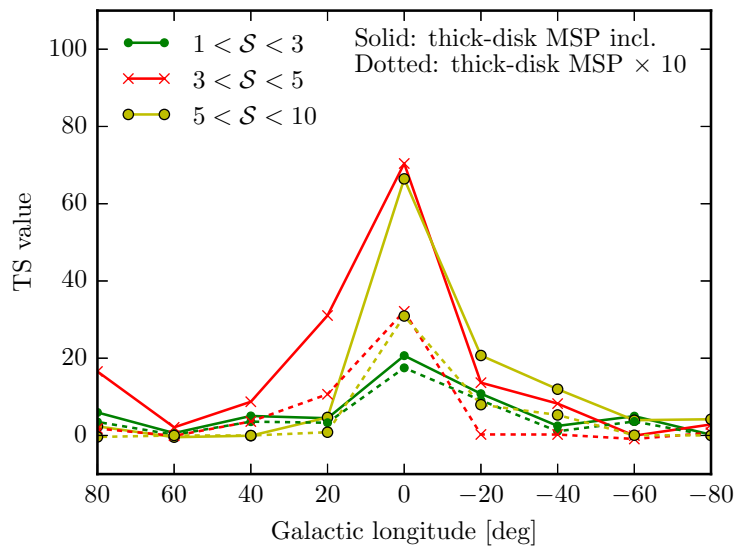}
  \end{center}
  \caption{Similar to Fig.~\ref{fig:TSs}, but including a thick-disk
    MSP population calibrated to local bright high-latitude MSPs as additional
    background.  Only for illustration, we also show the effect of a $10\times$
    more dense thick-disk population (which is in contradiction with local
    observations). Note that we unmask 3FGL as described in the main analysis
    when deriving the wavelet peaks.}
  \label{fig:TSsub}
\end{figure}

As argued above, the most relevant Galactic background in our ROI are expected
to be pulsars, and in particular the MSP thick-disk population that reaches up
to high latitudes.  We will now show that a thick-disk population of MSPs
(or other sources with a similar luminosity function) cannot be responsible for the observed signal.

\medskip

In most cases, the thick-disk population of MSPs is modeled as a cylindrically
symmetric exponential distribution, with a scale height in the range
0.5--$1\kpc$ and a scale radius of a few kpc, which is only poorly constrained
by data (see \textit{e.g.}~Ref.~\cite{Gregoire:2013yta}).  We will adopt here a
distribution with a scale height of $1\kpc$ and a scale radius of $5\kpc$,
which was previously used to argue against the MSP-origin of the \Fermi\ GeV
excess~\cite{Cholis:2014lta}. The distribution reads $n\propto
\exp\left(-R/R_\mathrm{s}\right)\exp\left(-\left|z\right| /
z_\mathrm{s}\right)$, with $R_{\rm s} = 5\mathrm{\,kpc}$ and $z_\mathrm{s} =
1\mathrm{\,kpc}$.  We will address below how the results change when other
parametrizations are adopted.

As \gr\ luminosity function, we adopt an inverse power-law with
$\Lmin=10^{31}\ergs$, $\Lmax=7\times10^{34}\ergs$ and index $\alpha=1.5$.  We
fix the overall normalization of the disk source density such that the number
of bright MSPs at high latitudes, $|b|>15^\circ$, is consistent with the number
of such MSPs listed in the 3FGL.  As flux threshold for bright MSPs we adopt a
flux that corresponds to a \gr\ luminosity of $10^{34}\ergs$ at $3\kpc$
distance ($9.2\times10^{-12}\ergs\rm cm^{-2}$ in our energy range).  We find 31
MSPs above that threshold flux and note that since the number of unassociated
high-latitude bright nonvariable sources with a curved enough spectrum is
small, this number cannot increase by more than 50\% when more unassociated
sources are identified as MSPs)~\cite{TheFermi-LAT:2015hja}.  For the present
scenario, we find that the total number of thick-disk sources with \gr\ 
luminosity above $10^{31}\ergs$ is $\sim30000$.

Within $2\kpc$ of the Galactic center, this thick-disk population predicts
around 1300 MSPs, which is more than an order of magnitude below the number
that we find in the best-fit scenario for the bulge population (around 35000
MSPs above $10^{31}\ergs$).  This implies, as already argued in
Ref.~\cite{Cholis:2014lta}, that a thick-disk population with the adopted
geometry cannot be responsible for the \Fermi\ GeV excess.  However, it also
trivially implies that the number of wavelet peaks caused by thick-disk sources
in the inner $2\kpc$ is about an order of magnitude below what is predicted by
our best-fit bulge population, and hence an order of magnitude below what is
actually observed.  This still leaves the possibility that thick-disk MSPs on
the line-of-sight towards the inner Galaxy, outside of the inner $\sim2\kpc$,
could affect our results.  We will discuss this next.

Within our ROI, the thick-disk population predicts 3.3 sources outside of the
inner $2\kpc$ with a flux in the range $(4.6$--$7.7)\times10^{-12}\ergs\rm
cm^{-2}$ (this corresponds roughly to $(4$--$7)\times10^{34}\ergs$ when
the sources are put at $8.5\kpc$ distance).  These sources could reasonably
contribute to wavelet peaks in the $3<\mathcal{S}<5$ range.  The actually
observed number of peaks in that range above the null hypothesis is about 35.
It is hence clear that foreground sources from the above thick-disk population
cannot cause the observed signal.  One might think of two ways around.

First, one could reduce the scale radius of the thick-disk population such that
the number of sources in the inner $2\kpc$ increases by a factor around ten
(scale radii around 1--$2\kpc$ could do the job).  This would give rise to a
wavelet signal similar to what is observed. However, such a population would
also predict a significant diffuse \gr\ emission similar to the level of
the \Fermi\ GeV excess, just with a morphology that is incompatible with the
observations.  A population with a scale radius of $1$--$2\kpc$ would indeed
commonly be referred to as bulge population.  Such a population would be very
similar to the bulge population that we put forward in the main part of the
paper as explanation for the \Fermi\ GeV excess, with the main difference being
that our population fits better the excess morphology.

Second, one could increase the number of MSPs in a ring-like region around the
Galactic bulge, excluding the inner $2\kpc$, such that these additional
ring-like distributed sources will enhance the number of foreground sources
without affecting the number of sources in the Galactic bulge.  In this case,
however, the wavelet signal should clearly be more extended along the Galactic
disk than what is shown in Fig.~\ref{fig:TSs}, since such a ring would not be
centrally peaked and extend to longitudes of at least $\sim25^\circ$. For
illustration, we here quote the relative number of wavelet peaks one expects in
the control regions along the disk and the main ROI produces by such a ring (1
kpc scale height, 5 kpc scale radius, the inner 2 kpc radius excluded): $\Delta
\ell = \left\{ \pm80, \pm60, \pm40, \pm20, 0\right\}$ and $N_\mathrm{peaks}
\propto  \left\{1.7, 2.4, 3.5, 4.9, 3.6\right\}$.  Moreover, in order to avoid
a conflict with the above calibration with bright high-latitude sources, the
ring should be further constrained to lie within $\lesssim5\kpc$, which however
would still leave a too flat central distribution of wavelet peaks.

\medskip

Finally, we show in Fig.~\ref{fig:TSsub} how the TS values are
affected if the thick-disk population (or a $10\times$ denser population) is added
as an additional background component.  For simplicity, we assume that the
thick-disk population causes deviations of the expectation values $\mu_{ij}$ in
Eq.~\eqref{eqn:likelihood} from the null hypothesis that are proportional to
the deviations caused by the best-fit bulge population.  We adjust the
normalization of these deviations such that the number of additionally
predicted $3<\mathcal{S}<5$ peaks in the main ROI is 3.3 (as motivated by the
above discussion).  We then add these thick-disk-induced deviations from the
null hypothesis as negative and positive contributions to the model predictions
in Eq.~\eqref{eqn:likelihood}, and repeat the CSP fit to inner Galaxy data.  We
repeat this procedure in all of the control ROIs used in Fig.~\ref{fig:TSs},
reweighing the thick-disk contribution properly at different Galactic
longitudes.  From Fig.~\ref{fig:TSsub} it is clear that only a thick disk ten
times denser than what is actually observed at higher latitudes could
significantly affect, although not completely remove, the excess of wavelet
peaks in the inner Galaxy.

\subsection{H. Further discussions}

\begin{figure}
  \begin{center}
        \includegraphics[width=0.55\linewidth]{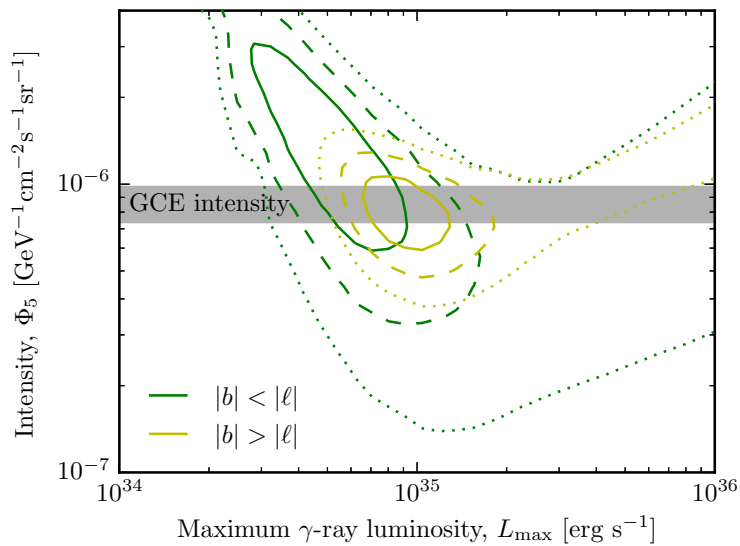}
  \end{center}
  \caption{Similar to Fig.~\ref{fig:limits} in the main text, but derived in
    redefined regions within our main ROI that are useful to check for the
    sphericity of the signal (we still mask $|b|<2^\circ$).  We find best-fit
    parameters for $\Lmax$ and $\Phi_5$ in the north/south and east/west
    regions that are consistent within one sigma.}
  \label{fig:sphericity}
\end{figure}

In order to test whether the spatial distribution of wavelet peaks in our
analysis is indeed compatible with a spherically symmetric distribution, we
re-binned the wavelet peaks into a north/south region, defined by
$12^\circ>|b|>\max(|\ell|,2^\circ)$, and an east/west region, defined by
$12^\circ>|\ell|>|b|>2^\circ$.  For each of the two regions, we derive the
best-fit values for $\Lmax$ and $\Phi_5$.  The results are shown in
Fig.~\ref{fig:sphericity}.  We find that the inferred parameters are consistent
within one sigma, with a slightly stronger signal in the north/south region.
This indicates that the excess of wavelet peaks, if interpreted in terms of a
bulge source population, is consistent with a spherical distribution of these
sources.

\medskip

Given that unresolved sources only add positively to the Galactic diffuse
emission, whereas a mismodeling of the gas could cause both positive and
negative variations, it is tempting to think that a detection of negative
wavelet peaks would disfavour an interpretation in terms of unresolved point
sources.  In fact, we do find a suppression of $-2<\mathcal{S}<-1$ peaks, and
an enhancement of $-4<\mathcal{S}<-3$ peaks in the inner ROI when searching for
negative instead of positive peaks.  But unfortunately, this cannot easily be
used to discriminate diffuse modeling artefacts (which, as we discussed above,
are anyway unlikely, as they should show up in the entire disk) from
sub-threshold point sources.

Maybe somewhat un-intuitively, negative wavelet peaks can indeed be generated
by a large number of weak, positive point sources.  This happens in the tails
of our simulated sources, where the wavelet transform becomes negative (this
effect is visible as rings around bright sources in Fig.~1).  We estimated the
expected number of \emph{negative} wavelet peaks for the best-fit scenario in
Fig.~3 by Monte Carlo simulations, and find results that are completely
consistent with the observed number of negative peaks.  However, given that the
number of positive and negative wavelet peaks are correlated (they are caused
by the same sources), one cannot easily use observations of negative wavelet
peaks to further constrain the model parameters.  This, and the fact that an
appropriate masking of 3FGL sources (including also the ring around each
source) reduces significantly the effective size of the ROI, make an efficient
use of negative peaks in our analysis difficult.  However, it is re-ensuring
that both the observed negative and positive wavelet peaks are consistent with the
respective predicted number of negative and positive wavelet peaks for the same
sub-threshold point source population.

\medskip

Finally, we briefly comment on the recent analysis of the Galactic center data
by the \Fermi-LAT collaboration \citep{TheFermi-LAT:2015kwa} and compare their
use of wavelets to ours.  Ref.~\citep{TheFermi-LAT:2015kwa} uses wavelets to
find seeds for the identification of point sources (also see
Ref.~\citep{Damiani:1997} for details about the adopted method), followed by 
standard maximum-likelihood fits for further source identification. This can
be potentially affected by interstellar emission modelling.
We instead study the statistics of
local-maxima in the wavelet-transformed sky map, which is largely background-model independent 
(although small scale fluctuations could in principle be relevant, as discussed above).  We find
good correspondence between our wavelet peaks and the 3FGL catalogue (see
Fig.~\ref{fig:transform}), which supports the validity of our approach.
However, we also find that this agreement and the quality of the wavelet
analysis in general critically depends on the adopted wavelet type and size.
All these points make it difficult to directly compare our results with those
of Ref.~\citep{TheFermi-LAT:2015kwa}.  However, we note that the Galactic disk,
where Ref.~\citep{TheFermi-LAT:2015kwa} finds that their identified sources most
strongly trace the edges of the interstellar emission and thus might
constitute false positives due to gas fluctuations (see their Fig.~8),
is masked in our analysis.

\end{document}